\documentclass[reprint,superscriptaddress,showpacs,amsmath,amssymb,aps]{revtex4-1}

\usepackage{wrapfig}
\usepackage{graphicx}  
\usepackage{dcolumn}   
\usepackage{bm}        
\usepackage{amssymb}   
\usepackage{mathrsfs}
\usepackage{amsthm}
\usepackage{mathtools}
\usepackage{framed}
\usepackage[colorlinks=True,urlcolor=blue,linkcolor=blue,citecolor=blue]{hyperref}
\usepackage{xcolor}
\usepackage{multirow}
\usepackage{setspace}
\usepackage{CJKutf8}
\usepackage{color,soul}

\usepackage[utf8]{inputenc}
\usepackage[T1]{fontenc}
\usepackage{lmodern}
\usepackage[toc,page]{appendix}
\usepackage{tikz}
\usetikzlibrary{arrows,arrows.meta,decorations.markings,decorations.pathreplacing,calc}
\tikzset{->-/.style={decoration={markings,mark=at position #1 with {\arrow{Stealth}}},postaction={decorate}},->-/.default=0.6}

\theoremstyle{definition}
\newtheorem{defn}{Definition}
\newtheorem{thm}[defn]{Theorem}
\newtheorem{rem}[defn]{Remark}
\newtheorem{expl}[defn]{Example}

\def\t#1{\widetilde{#1}}

\def\dd{\mathrm{d}}

\newcommand{\be}{\begin{equation}}
\newcommand{\ee}{\end{equation}}
\newcommand{\Z}{\mathcal{Z}}

\DeclareMathOperator{\id}{id}

\begin{document}

\begin{CJK*}{UTF8}{}

\title{Classification of topological phases in one dimensional interacting non-Hermitian systems and emergent unitarity }

\author{Wenjie Xi}
\affiliation{Shenzhen Key Laboratory of Advanced Quantum Functional Materials and Devices, Southern University of Science and Technology, Shenzhen 518055, China}
\affiliation{Institute for Quantum Science and Engineering and Department of Physics, Southern University of Science and Technology, Shenzhen 518055, China}

\author{Zhi-Hao Zhang}
\affiliation{School of Mathematical Sciences, University of Science and Technology of China, Hefei 230026, China}
\affiliation{Institute for Quantum Science and Engineering and Department of Physics, Southern University of Science and Technology, Shenzhen 518055, China}

\author{Zheng-Cheng Gu}
\email{zcgu@phy.cuhk.edu.hk}
\affiliation{Department of Physics, The Chinese Uinversity of Hong Kong,  Hong Kong, China}

\author{Wei-Qiang Chen}
\email{chenwq@sustech.edu.cn}
\affiliation{Shenzhen Key Laboratory of Advanced Quantum Functional Materials and Devices, Southern University of Science and Technology, Shenzhen 518055, China}
\affiliation{Institute for Quantum Science and Engineering and Department of Physics, Southern University of Science and Technology, Shenzhen 518055, China}


\begin{abstract}
Topological phases in non-Hermitian systems have become fascinating subjects recently. In this paper, we attempt to classify topological phases in 1D interacting non-Hermitian systems. We begin with the non-Hermitian generalization of the Su-Schrieffer-Heeger (SSH) model and discuss its many-body topological Berry phase, which is well defined for all interacting quasi-Hermitian systems (non-Hermitian systems that have real energy spectrum).
We then demonstrate that the classification of topological phases for quasi-Hermitian systems is exactly the same as their Hermitian counterparts.
Finally, we construct the fixed point partition function for generic 1D interacting non-Hermitian local systems and find that the fixed point partition function still has a one-to-one correspondence to their Hermitian counterparts. Thus, we conclude that the classification of topological phases for generic 1D interacting non-Hermitian systems is still exactly the same as Hermitian systems.

\begin{center}
\end{center}

\end{abstract}

\maketitle
\end{CJK*}

\section{Introduction}

Recently, the topological properties of non-Hermitian Hamiltonians have drawn much attention both experimentally~\cite{xiao2019,Kawakami2017,Merchdeh2018,Alexander2015,Zhou2017Observation,Yi2021} and theoretically~\cite{Wang2018,Fu2017,TonyLee2016,Elganainy2018,Franco2017,Franco2019_1,Shuichi2019,Masahito2018,Ronny2019,Takehito2019,Duan2017,Notomi2018,Hui2018,Nicolas2019,chang2019,Vatsal2019,yoshida2019nonhermitian,SongZ2019}. Studies have been focused intensively on topological phases in non-Hermitian free fermion systems~\cite{Yeon2019,Gong2018,Kawabata_2019} and the phase transitions among them~\cite{Fu2017,Rudner2009,Longhi2019}. Unfortunately, a general classification scheme of topological phases in interacting non-Hermitian systems is still lacking.
In principle, the concept of the entanglement pattern (with or without global symmetries) can still be used to define and classify topological phases for non-Hermitian systems and characterize their ground state properties, although some technical details such as local unitary (LU) transformations need to be modified to fit the non-unitary time evolution of non-Hermitian systems. Meanwhile, as a fundamental property of ordinary quantum mechanics, unitarity plays an essential role in modern physics, and it will be of great interest to investigate how unitarity can emerge from an underlying non-unitary system.

In this paper, we attempt to provide a complete classification of topological phases in $1$D interacting non-Hermitian systems.
In Section~\ref{sec:ssh}, we first demonstrate that the classification of topological phases for quasi-Hermitian systems with real energy spectrum is the same as their Hermitian counterparts with a simple example.
This follows from the fact that these kinds of non-Hermitian Hamiltonians can always be mapped to their unique Hermitian counterparts via local similarity transformation. Thus, they have the same topological invariants as their Hermitian counterparts, while the edge states can also be mapped onto those of the Hermitian case with a local similarity transformation.
Physically, the topological Berry phase of quasi-Hermitian systems can be regarded as a $\mathbb{C^\times}$-valued phase factor instead of the $\mathrm U(1)$ valued phase factor in the usual Hermitian systems under adiabatic evolution.
In Section \ref{sec:classification_bosonic}, it is shown that for 1D quasi-Hermitian systems, the unitarity condition will naturally emerge in the fixed point topological invariant partition function and the classification of bosonic symmetry-protected topological (SPT) phases is exactly the same as their Hermtian counterparts, which are classified by second group cohomology $H^2(G,\mathrm U(1)_T)$~\cite{Wen2011,Gu2011,fidkowski11,schuch11, chenScience2012, chen13}.
On the other hand, as $1$D local fermionic systems can always be mapped to $1$D local bosonic systems, the above conclusion for SPT phase also holds for 1D quasi-Hermitian interacting fermion systems. As Kitaev's Majorana chain is the only intrinsic fermionic topological phase, we will study this specific case for quasi-Hermitian systems in the Supplementary materials.

Given the above, we study non-Hermitian systems with complex energy spectrum in Section~\ref{sec:classification_general}. In this case, topological invariants are not always well defined, as the ground state is allowed to bypass an excited state without level crossing. However, it is elaborated in Section~\ref{sequence} and \ref{sec:tqft} that topological quantum field theory (TQFT) still captures all possible topological phases for 1D non-Hermitian quantum systems (including those non-Hermitian systems with complex eigenvalues). These phases have a one-to-one correspondence to their Hermitian counterparts. For bosonic SPT phases, this is simply because $H^2(G,\mathrm U(1)_T)$ is isomorphic with $H^2(G,\mathbb{C}^\times_T)$, thus the classification of their topological phases is exactly the same.

\begin{figure}
\includegraphics[width=9cm]{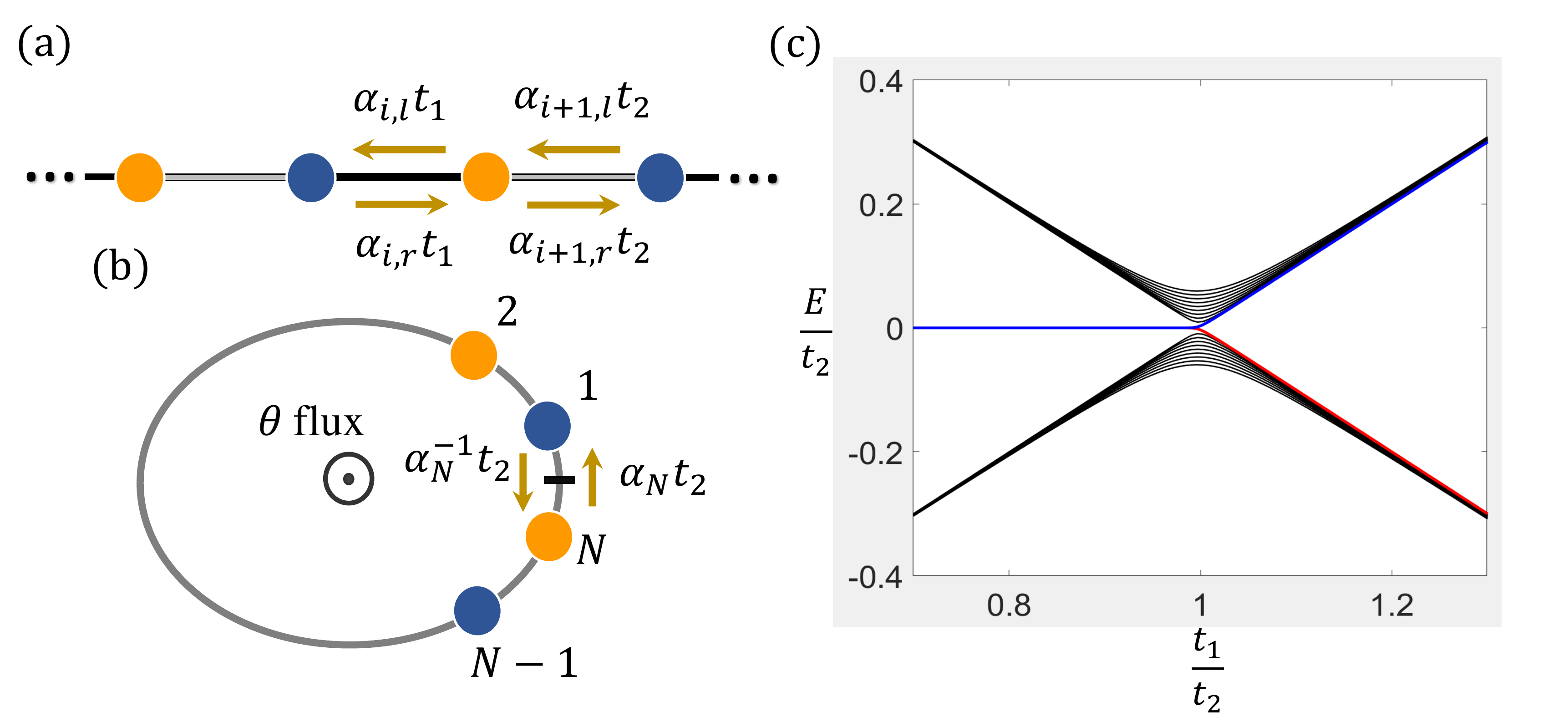}
\caption{\label{fig:fig1}
(a) Illustration of a non-Hermitian non-interacting Su-Schrieffer-Heeger (SSH) model of the nearest neighbor hopping of spinless electrons on a one-dimensional chain with two atoms per unit cell.  (b) Schematic picture of the non-Hermitian SSH model on a ring with a twisted boundary condition and an inserted $\theta$ flux. (c) Illustration of single-particle energy spectrum of both Hermitian and quasi-Hermitian SSH model under OBC with $N=1000$. The two bulk states marked by red and blue become zero energy edge states as the value of $t_1/t_2$ decreases. }
\end{figure}

\section{A simple example: Quasi-Hermitian Su-Schrieffer-Heeger (SSH) model with interactions} \label{sec:ssh}
Without loss of generality, we take the SSH model, which is well studied for both non-interacting~\cite{Wunner2017ssh,Gloria2019,Weber_2015} and interacting~\cite{Victor2012,Yoshihito2019,Marques_2018,Dias2017} cases, as a simple example.
We begin with a 1D non-Hermitian non-interacting SSH model with nearest neighbor hopping for spinless fermion system,
as shown in Fig.~\ref{fig:fig1}a, with the Hamiltonian:
\begin{align}
\label{eq:eqn2}
{H}_{\rm SSH}=& \sum_{i=2n-1} \bigl(  \alpha_{i,r} t_1{c}^\dagger_{i+1}{c}_{i}+\alpha_{i,l}t_1{c}^\dagger_{i}{c}_{i+1} \bigr) \nonumber\\
& + \sum_{i=2n} \bigl(\alpha_{i, r}t_2{c}^\dagger_{i+1} {c}_{i}+\alpha_{i, l}t_2{c}^\dagger_{i}{c}_{i+1} \bigr),
\end{align}
where $c_{i}$ is the annihilation operator of electron on site $i$, $t_1$ and $t_2$ are real numbers related to the hopping integrals of electrons between nearest sites in the same unit cell and in a different unit cell respectively, $\alpha_{i, l(r)}$ are nonzero real numbers.
If we take $\alpha_{i,l}=\alpha_{i,r}=1$, it reduces to the well-known Hermitian SSH model.  For a half-filled Hermitian SSH model, there is a topologically nontrivial phase at $t_1<t_2 $ and a phase transition at $t_1 = t_2 $ to a topologically trivial phase.  The topological properties of such a phase can be described by the topological invariants, i.e., the winding number in $k$-space, of the bulk state with periodic boundary conditions (PBC) or the zero-energy state at the edge with open boundary conditions (OBC).  This is the so-called bulk-boundary correspondence in free fermion systems. The topologically nontrivial ground state is protected by charge conservation and anti-unitary chiral symmetry $S$, defined as:
\begin{align}
\label{symmetry}
S{c}_{i}S^{-1}= (-1)^i {c}^\dagger_{i},\quad
S{c}^\dagger_i S^{-1}=(-1)^i {c}_{i} \quad S \mathrm i S^{-1}=-\mathrm i.
\end{align}
Note that although we need translational symmetry to calculate the winding number in $k$-space, the topological phase is not protected by the translational symmetry.

Then we move to the non-Hermitian SSH model and check whether there is still a nontrivial SPT phase or not.  First, we consider a special case with $\alpha_{i,l}= \alpha^{-1}_{i,r} = \alpha_{i}$, which can be mapped to the Hermitian SSH model via a local similarity transformation:
\begin{eqnarray}
\label{eq:similarity}
c_i \rightarrow \left(\prod_{j=1}^{i-1}\alpha^{-1}_j\right) {c}_{i}, \quad c^\dagger_i \rightarrow \left(\prod_{j=1}^{i-1}\alpha_j\right) {c}_{i}^\dagger.
\end{eqnarray}
This means the non-Hermitian model has a real spectrum even though its Hamiltonian is non-Hermitian.  In the following, we refer the non-Hermitian Hamiltonians with real spectrum as quasi-Hermitian Hamiltonians (the term ``qusi-Hermitian'' is borrowed from Ref. \cite{qHqm}).
In fact, the quasi-Hermitian model \eqref{eq:eqn2} has exactly the same spectrum as the Hermitian SSH model as shown in Fig.~\ref{fig:fig1}c, which means that there are zero-energy edge states at $t_1 < t_2$ but no zero-energy edge state at $t_1 > t_2$.  However, this does not mean that it is a topologically nontrivial phase.  We must prove that the zero-energy edge states are related to some topological properties of the bulk and protected by certain symmetries, especially for interacting systems.

To detect the topological properties of the bulk, we need to connect the two ends of the chain to form a ring, as shown in Fig.~\ref{fig:fig1}b, and calculate some topological invariants of the system on this ring.  In the Hermitian case, one uses the PBC, which corresponds to connecting the two ends with $t_2{c}^\dagger_{N} {c}_{1} + t_2{c}^\dagger_{1}{c}_{N}$, and calculating the winding number in $k$-space.  However, in the most general quasi-Hermitian cases, the system does not have translational symmetry, so one cannot do calculations in $k$-space.  Instead, we must use the twist boundary condition (TBC)~\cite{Wu2006}, introduced by Wu et al.~\cite{Wu1985} to study quantum Hall states, and to calculate the topological invariant of the ground state.  The basic idea is to introduce an additional phase factor $e^{i\theta}$ in the boundary conditions, i.e., $e^{i \theta} t_2{c}^\dagger_{N} {c}_{1} + e^{-i \theta} t_2{c}^\dagger_{1}{c}_{N}$,  which corresponds to inserting a $\theta$ flux in the center of the ring as shown in Fig.~\ref{fig:fig1}b.  Obviously, the ground state wave function depends on the phase $\theta$.
By further assuming that the many-body ground state of the system is separated from the excited states by a finite gap for all values of the twisted phase $\theta$, one can define a topological invariant by the total flux of the Berry-phase gauge field associated with the ground state over the $\theta$-space, i.e.,
\begin{eqnarray}
\label{eq:eqn3}
C=\frac{i}{\pi}\int_0^{2\pi}\langle\varphi_G(\theta)\vert \frac{\partial}{\partial \theta} \vert \phi_G(\theta)\rangle \mathrm{d}\theta,
\end{eqnarray}
where $\langle\varphi_G(\theta)\vert(\vert \phi_G(\theta)\rangle)$ is the left (right) many-body ground state of the non-Hermitian Hamiltonian.  The advantage of the TBC is that it can handle general cases with interactions and without translational symmetry.

A simple calculation shows that the resultant $C$ gives the correct bulk-boundary correspondence in the Hermitian case.  However, such correspondence is absent in the quasi-Hermitian case.  This can be understood from the similarity transformation Eq.~\eqref{eq:similarity}.  After the transformation, the quasi-Hermitian SSH model with PBC is mapped to a Hermitian SSH model with a boundary condition $\gamma t_2 \tilde{c}^\dagger_{1} \tilde{c}_{N} + \gamma^{-1} t_2 \tilde{c}^\dagger_{N} \tilde{c}_{1}$ with $\gamma = \prod_{i=1}^{N-1} \alpha_{i}$.  Such a boundary condition breaks the chiral symmetry, and thus the bulk-boundary correspondence of the SPT state.  However, if we consider a different boundary condition:
\begin{align}
 \alpha_{N} t_2{c}^\dagger_{1} {c}_{N} + \alpha^{-1}_{N}t_2 {c}^\dagger_{N} {c}_{1}, \label{boundary}
\end{align}
with $\alpha_N = \gamma^{-1} = \prod_{i=1}^{N-1} \alpha_i^{-1}$, the corresponding Hermitian Hamiltonian after the mapping will be the Hermitian SSH model with PBC, and one should have the correct bulk-boundary correspondence.  This has been confirmed by our TBC calculations, which show the winding number to be 1 for $t_1 < t_2$ and 0 for $t_1 > t_2$ as long as all of the $\alpha$s satisfy $\prod_{i=1}^{N} \alpha_{i} = 1$.  For the case $\alpha_{1}=\alpha_{2}=\cdots=\alpha_{N-1}=\alpha$,
one should impose a boundary condition with $\alpha_N = \alpha^{-(N-1)}$. Note that the similarity transformation Eq.~\eqref{eq:similarity} indicates an imaginary term in $k$ after Fourier transformation of $c_i$.  This provides an explanation of the failure of the conventional winding number in $k$-space in the non-Hermitian SSH model and the success of the winding number in a complex $k$-space in Ref.~\cite{Wang2018}.

\begin{figure}
\includegraphics[width=8.5cm]{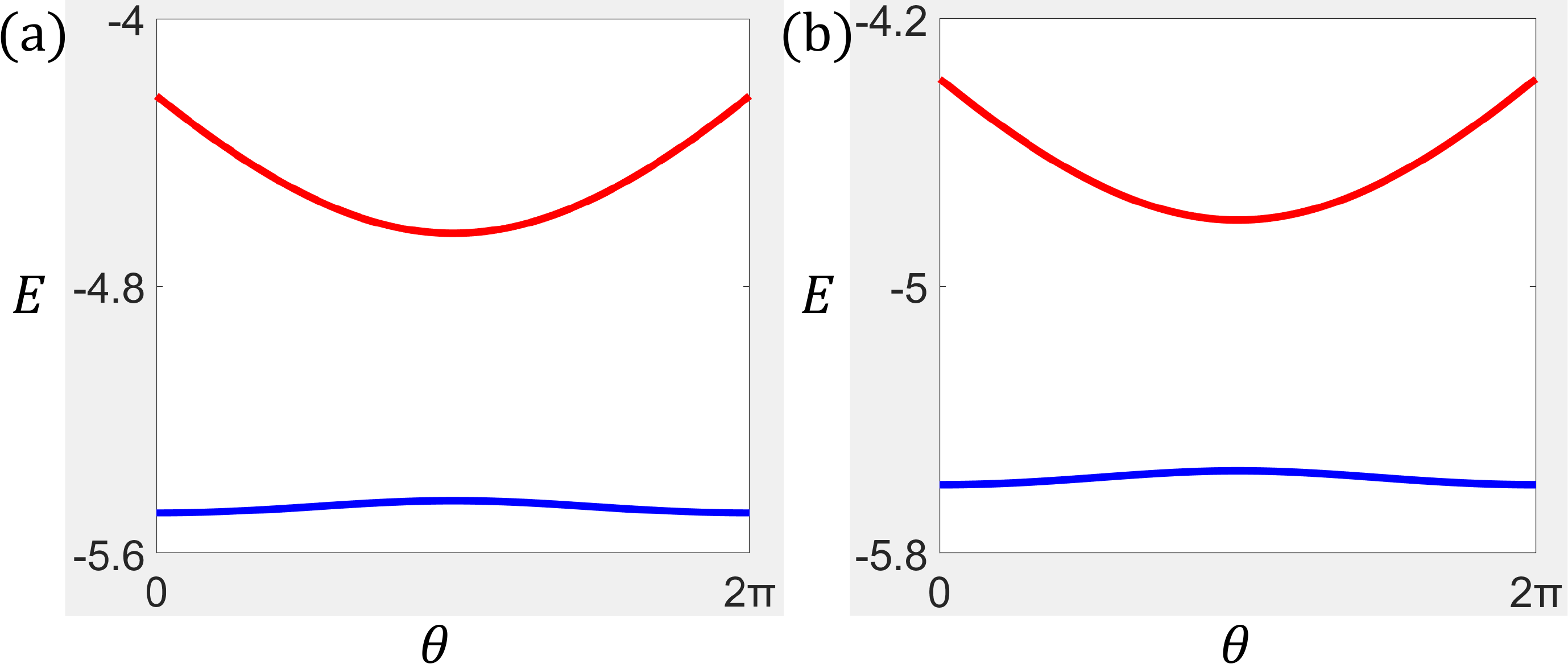}
\caption{\label{fig:fig2}
Illustration of many-body energy spectrum of ground state and first excited state of Hamiltonian \eqref{H_inter} with twist phase $\theta$ and $t_1=0.6$, $t_2=1$, $N = 10$.  (a) The non-interacting quasi-Hermitian case with $U=0$ and $\mu=0$.  (b) The interacting quasi-Hermitian case with $U = 0.1$ and $\mu = 0$.  In (a) and (b), the spectrum is gapped for all values of $\theta$, and thus the topological invariant is still well defined.
}
\vspace{-4pt}
\end{figure}

To check whether the topological phase is protected, we introduce a small imaginary chemical potential difference and a small Coulomb repulsion between electrons in the same unit cell, and the Hamiltonian becomes:
\begin{align}
\label{H_inter}
{H}=& H_{SSH} + i \mu \sum_{j} n_{2j-1} + U \sum_j ({n}_{2j-1}-\frac{1}{2})( {n}_{2j}-\frac{1}{2}).
\end{align}
We first consider a simple case with $U = 0.1$ and $\mu = 0$, where the interaction terms after the similarity transformation Eq.~\eqref{eq:similarity} also respect the chiral symmetry Eq. ~\eqref{symmetry} of the Hermitian SSH model.
The winding number calculated with TBC depicted in Fig.~\ref{fig:fig2}b demonstrates that the ground state remains topologically nontrivial at $t_1 < t_2$.  It is reasonable to think that the topologically nontrivial phase discovered above is protected by a quasi-Hermitian version of chiral symmetry $\t S$:
\begin{eqnarray}
\label{eq:eqn4}
\t S{c}_{i}\t S^{-1}&= (-1)^i \left( \prod_{j=1}^{i-1}\alpha_j^{-2} \right) {c}^\dagger_{i},\quad
\t S{c}^\dagger_i \t S^{-1}=(-1)^i  \left( \prod_{j=1}^{i-1}\alpha_j^2 \right) {c}_{i} \nonumber\\
\t S \mathrm i \t S^{-1}&=-\mathrm i,
\end{eqnarray}
which is related to $S$ by the similarity transformation Eq.~\eqref{eq:similarity}. It should be noticed the action of symmetry $\t S$ for non-Hermitian systems can be a similarity transformation.
Again, according to the similarity transformation Eq.~\eqref{eq:similarity}, the phase should belong to the same topological phase as its corresponding phase of the Hermitian SSH model.
There is a more profound and general understanding of the similarity transformation Eq.~\eqref{eq:similarity}.  In quantum mechanics, two wave functions $\psi$ and $\phi$ that differ by a nonzero complex factor, i.e., $\phi = z \psi$ with $z \neq 0 $, correspond to the same physical state.  A theory should be invariant under local gauge transformation ${c}_i \rightarrow e^{-i\theta_i} {c}_{i},~ {c}^\dagger_i \rightarrow e^{i\theta_i} {c}_{i}^\dagger$.  If $\theta_i$s are real numbers, they are just the $\mathrm U(1)$ gauge choice of the local basis. However, for quasi-Hermitian systems, we can consider the most general $\mathbb{C^\times}$ gauge choice with complex $\theta_i$'s (here $\mathbb{C}^\times = \mathbb{C} \setminus \{0\}$ is the group of non-zero complex numbers),
and the similarity transformation Eq.~\eqref{eq:similarity} is exactly a local $\mathbb{C}^\times$ gauge transformation.
Thus, the corresponding topological Berry phase, which arises from adiabatic evolution, should be described by a $\mathbb{C}^\times$-valued gauge field for generic quasi-Hermitian systems.

\section{Classification of topological phases in 1D quasi-Hermitian systems} \label{sec:classification_bosonic}
Now we move the discussion to generic interacting quasi-Hermitian systems. For $1$D Hermitian bosonic systems without any symmetry, it is well known that any gapped quantum state can always connect to a trivial product state without phase transition\cite{frank2005,Wen2011}.  Obviously, such a statement still holds for 1D non-Hermitian bosonic systems, and SPT phases are still the only possible topological phases in bosonic case.  Fermionic systems is in general much more complicated.  However,  it has been shown that the classification of 1D Hermitian fermionic SPT phases can be obtained from that for bosonic systems via the Jordan-Wigner transformation (i.e., bosonization) \cite{Wen2011,Jordan1993}, and this should still hold in quasi-Hermitian systems.  Thus we only need to focus on the classification of bosonic SPT phases in quasi-Hermitian systems in this section.  In Hermitian cases, the bosonic SPT phases in $n$ dimension have been successfully classified by classifying the fixed point partition functions.  Mathematically, the classification is given by the group cohomology $H^n(G,\mathrm U(1)_T)$ \cite{chenScience2012}.  In this section, we will generalize this approach to quasi-Hermitian systems.

The topologically invariant fixed point partition function for bosonic SPT phases protected by a finite symmetry group $G$ in quasi-Hermitian systems (defined on an arbitrary branched triangulation of a 2D manifold) can be written as
\begin{align}
Z_f = &\frac{1}{|G|^{N_v}}\sum_{\{g_i\}}\prod_{\rm{triangular}} \nu_{2}^{s_{ijk}}(g_i,g_j,g_k), \label{action}
\end{align}
where $|G|$ is the order of the group, with $N_v$ the number of total vertices, and $s_{ijk}=\pm$ is determined by the orientation of the corresponding triangulation.
Eq.~\eqref{action} has exactly the same form as the fixed point partition function of 1+1D Hermitian systems except that $\nu_2^{\pm}(g_i,g_j,g_k) \in \mathbb{C}^\times$ instead of $U(1)$ for quasi-Hermitian systems.
$\nu_2^{\pm}(g_i,g_j,g_k)$ is a function of group element $g_i, g_j, g_k$, which can be naturally regarded as the $\mathbb{C}^\times$-valued symmetric topological Berry phase term for quasi-Hermitian systems.
 $\nu_2^{\pm}(gg_i,gg_j,gg_k)=\nu_2^{\pm}(g_i,g_j,g_k)$ leaving the partition function invariant, which implies the partition function is invariant under symmetry operation.
Moreover, we can further impose the following condition:
\begin{align}
[\nu^{+}_2(g_i,g_j,g_k )]^*=
\nu^{-}_2(g_i,g_j,g_k )\label{conjugate}.
\end{align}
This condition holds because for quasi-Hermitian systems with real energy spectrum, the time reversal symmetry can always be realized by the complex conjugate operation. Alternatively, time reversal can also be defined as the reversing of branching arrows, which naturally reverses the time ordering and orientation for a given triangulation.

\begin{figure}
\includegraphics[width=8.5cm]{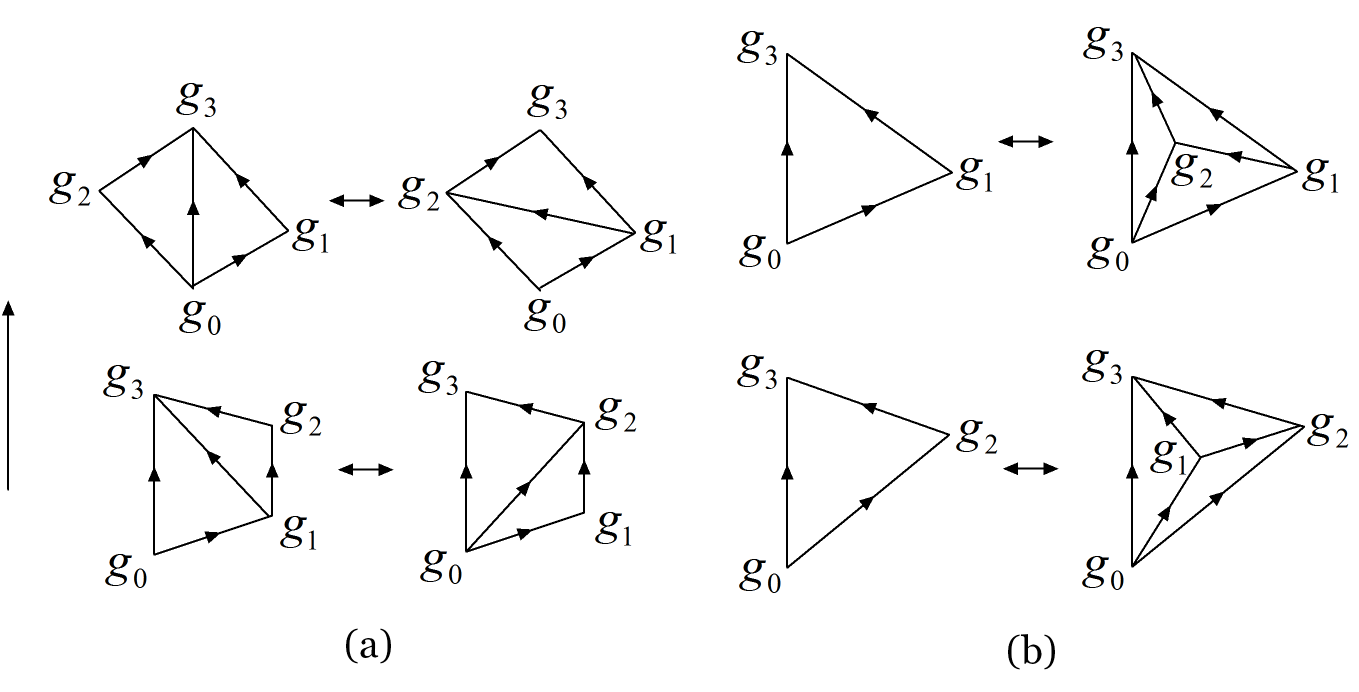}
\vspace{-4pt}
\caption{\label{fig:fig3}
2D Pachner moves (re-triangulations) with time ordering}
\vspace{-4pt}
\end{figure}

Furthermore, as a topologically invariant partition function, it must be invariant under all possible Pachner moves (re-triangulations) as depicted in Fig.~\ref{fig:fig3} for arbitrary branched triangulation.  For the $2 \leftrightarrow 2$ moves, we have:
\begin{align}
\nu^+_2(g_0,g_1,g_3)\nu^-_2(g_0,g_2,g_3)=&
\nu^-_2(g_1,g_2,g_3)\nu^+_2(g_0,g_1,g_2),\nonumber\\
\nu^+_2(g_1,g_2,g_3)\nu^+_2(g_0,g_1,g_3)=&
\nu^+_2(g_0,g_2,g_3)\nu^+_2(g_0,g_1,g_2).
\label{G1D22move}
\end{align}
Similarly, for the $1\leftrightarrow 3$ moves, we have:
\begin{align}
\nu^+_2(g_0,g_1,g_3)=&
\nu^-_2(g_1,g_2,g_3) \nu^+_2(g_0,g_2,g_3 )\nu^+_2(g_0,g_1,g_2 ),\nonumber\\
\nu^+_2(g_0,g_2,g_3)=& \nu^+_2(g_1,g_2,g_3) \nu^+_2(g_0,g_1,g_3)\nu^-_2(g_0,g_1,g_2).
\label{G1D13move}
\end{align}
Physically, the two Pachner moves correspond to retriangulation and coarse graining of the partition function.
All of the above four equations form a consistent algebra which is also satisfied by fixed point partition function of Hermitian systems, and they lead to the unitarity condition for $\nu_2^{\pm}$:
\begin{align}
\nu^{+}_2(g_i,g_j,g_k )
\nu^{-}_2(g_i,g_j,g_k )=1,\label{Tcondition}
\end{align}
which further unifies the above four equations into the well-known 2-cocycle equation of $\nu_2^+$:
\begin{align}
\nu^+_2(g_1,g_2,g_3)\nu^+_2(g_0,g_1,g_3)=
\nu^+_2(g_0,g_2,g_3)\nu^+_2(g_0,g_1,g_2).\label{cocycle}
\end{align}
Thus, we conclude that for quasi-Hermitian systems, SPT phases are still classified by $H^2(G,\mathrm U(1))$ and unitarity emerges for the fixed point partition function. The above results can also be generalized into anti-unitary symmetry cases straightforwardly.

As a summary, we have shown that the classification of 1D quasi-Hermitian bosonic SPT phases is exactly the same as the one of Hermitian systems.  In fact, the emergence of unitarity discussed above should also apply to all the quasi-Hermitian bosonic SPT phases in higher dimensions.  Moreover, because 1D local fermionic systems can always be mapped onto local bosonic systems as discussed in the beginning of this section, the above claim also holds for classifying SPT phases in 1D quasi-Hermitian fermionic systems. Besides the SPT phases, there is a so-called intrinsic topological phase in a 1D fermionic system, which is the Kitaev's Majorana chain. The above derivation of emergent unitarity is still correct for this case, see the Supplementary materials for more details.

\section{Topological phases in generic 1D non-Hermitian systems} \label{sec:classification_general}

\subsection{Non-Hermitian SSH model with complex energy spectrum}

Finally, we consider the most general non-Hermitian case, where the eigenvalues of the Hamiltonian can be complex. The ground state is decided by considering the path integral in Euclidean space.  At a large time scale, only state with lowest real part energy survives. Thus, we define the state with lowest real part energy as the ground state. The imaginary part energy of the ground state only contribute an oscillation of state. Again, we use the SSH model to illustrate the basic idea.

\begin{figure}
\includegraphics[width=8.5cm]{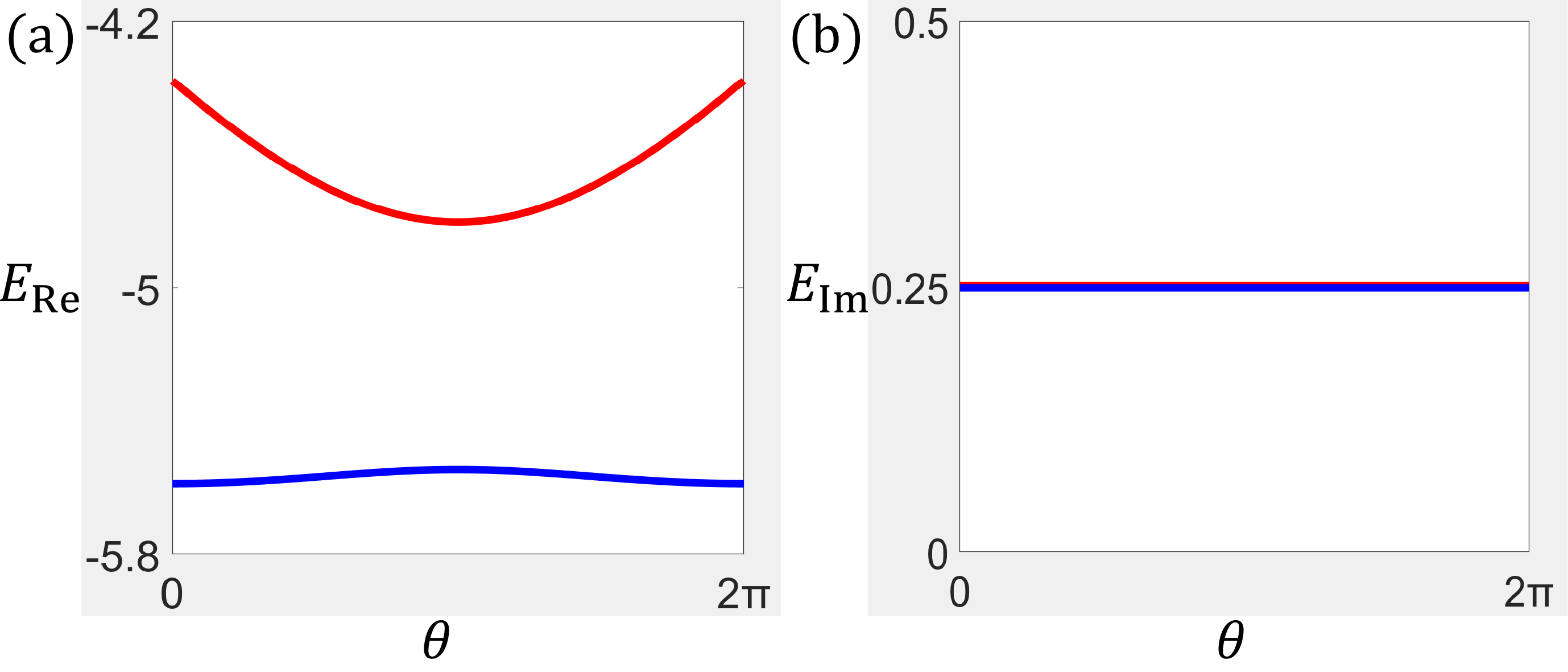}
\caption{\label{fig:fig5}
Illustration of many-body energy spectrum of ground state and first excited state of Hamiltonian \eqref{H_inter} with twist phase $\theta$ and $t_1=0.6$, $t_2=1$, $N = 10$. The real and imaginary part of the many-body energy spectrum in generic non-Hermitian case with $U=0.1$ and  $\mu=0.1$ are drawn in (a) and (b) respectively. It is obvious that the real part of the gap between ground state and first excited state shown in (a) is much bigger than the imaginary part of the ground state energy shown in (b).  Thus we can still have a well-defined topological invariant.}
\vspace{-4pt}
\end{figure}

In Fig.~\ref{fig:fig5}a and b, we show the real and imaginary parts of the spectrum of the model Eq.~\eqref{H_inter} with $\alpha_{i,l}=\alpha_{i,r}=1$, $t_1=0.6$, $t_2=1$, $\mu=0.1$ and $U = 0.1$ at various twist angles $\theta$.
Although defining symmetry is subtler for systems with complex eigenvalues, in this specific case the chiral symmetry can still be defined as Eq.~\eqref{symmetry}.
However, when we take $\theta$ from $0$ to $2\pi$, a generic state might bypass another state without level crossing,
and one may not be able to calculate the topological invariant $C$ for that state.  This happens only when the energy spectrum is complex.

Fortunately, the ground state, which is considered to have the lowest real part energy, still has a well-defined $C$ when $t_1 / t_2$ is away from 1 in this system.  It is not surprising because the imaginary chemical potential can be regarded as a perturbation in this case. From another perspective, as the gap between ground state and excited state is much larger than the imaginary chemical potential for any twist angle $\theta$, we shall not expect the imaginary component of the ground state energy raised from the imaginary chemical potential will change band dispersion dramatically. The corresponding numerical result shows that we still have $C = 1$ for $t_1 < t_2$ and $C = 0$ for $t_1 > t_2$.  For $t_1 \simeq t_2$, the gap is much smaller so that we cannot neglect the modification on band dispersion caused by the imaginary chemical potential.  The situation near the critical point is quite subtle and complicated. The ground state may bypass another state without level crossing and thus does not have a well defined $C$.
Nevertheless, our results still indicate a critical point at $t_1 \simeq t_2$. This suggests that there are still two SPT phases in this non-Hermitian case with a complex energy spectrum and they belong to the same fixed points as the Hermitian case.

The above example suggests that the emergent unitarity, i.e., a non-Hermitian SPT phase corresponding to a fixed point of the Hermitian case, still holds for 1+1D generic non-Hermitian systems. Whether the above statement holds for the most generic 1+1D non-Hermitian SPT phases can be studied by considering its fixed point theory, which will be studied in full details below.

\subsection{Classification of SPT phases in generic 1D non-Hermitian systems}
\label{sequence}

In fact, the topological invariant fixed point partition function Eq. \eqref{action} can also be defined for generic 1D non-Hermitian bosonic systems and we only need to remove the quasi-Hermitian condition Eq. (\ref{conjugate}). Similar to the quasi-Hermitian case, we can also derive the 2-cocycle condition Eq. \eqref{cocycle} by considering all possible Pachner moves. However, in this case, $\nu_2^+$ is $\mathbb{C}^\times_T$-valued and the classification of topological phases in generic 1D non-Hermitian bosonic systems is given by $H^2(G,\mathbb{C}^\times_T)$. Interestingly, mathematically it turns out that for any finite group or compact Lie group $G$, the natural inclusion $\mathrm U(1)_T \hookrightarrow \mathbb C^\times_T$ induces isomorphisms $H^i(G;\mathrm U(1)_T) \cong H^i(G;\mathbb C^\times_T)$ for all $i > 0$, i.e., the classification of the non-Hermitian systems is still the same as the corresponding Hermitian systems. The details are given below.

\medskip
At first, we consider a finite group $G$ corresponding to only unitary symmetries. Since all $\mathrm U(1)$-valued cochains are naturally $\mathbb C^\times$-valued cochains, the inclusion $\mathrm U(1) \hookrightarrow \mathbb C^\times$ induces homomorphisms $H^i(G,\mathrm U(1)) \to H^i(G,\mathbb C^\times)$ for all $i > 0$. It may be shown by direct calculation that a $\mathbb C^\times$-valued cocycle is automatically valued in $\mathrm U(1)$. However, we prove it in a more abstract way here.

Since $G$ represents unitary symmetries, it acts trivially on coefficient groups $\mathrm U(1)$ and $\mathbb C^\times$. It is known that the short exact sequence of (multiplicative) abelian groups with the trivial $G$-action
\begin{eqnarray}
1 \longrightarrow \mathrm U(1) \longrightarrow \mathbb C^\times \longrightarrow \mathbb R_+ \longrightarrow 1,
\end{eqnarray}
where $\mathbb R_+$ denotes the multiplicative group of positive real numbers, induces a long exact sequence of cohomology groups
\begin{eqnarray}
&& 1 \longrightarrow H^0(G,\mathrm U(1))\longrightarrow H^0(G,\mathbb C^\times) \longrightarrow H^0(G,\mathbb R_+)\nonumber \\
&& \phantom{1} \longrightarrow H^1(G,\mathrm U(1))\longrightarrow H^1(G,\mathbb C^\times) \longrightarrow H^1(G,\mathbb R_+)\nonumber \\
&& \phantom{1} \longrightarrow H^2(G,\mathrm U(1))\longrightarrow H^2(G,\mathbb C^\times) \longrightarrow \cdots\cdots
\end{eqnarray}
Since all cohomology groups $H^i(G,\mathbb R_+)$ are trivial for $i > 0$, and $H^0(G,A) = A$ for all $A$ with the trivial $G$-action, we get short exact sequences
\begin{eqnarray}
1 \longrightarrow H^i(G,\mathrm U(1)) \longrightarrow H^i(G,\mathbb C^\times) \longrightarrow 1
\end{eqnarray}
for all $i > 0$, which imply that $H^i(G,\mathrm U(1)) \cong H^i(G,\mathbb C^\times)$ for all $i > 0$.

If $G$ contains anti-unitary symmetries, for example the time-reversal symmetry $T$, then the actions of $G$ on coefficient groups $\mathrm U(1)$ and $\mathbb C^\times$ are non-trivial: if $T \in G$ is anti-unitary, we have $T \mathrm i T^{-1} = -\mathrm i$, i.e., $T$ acts by the complex conjugate. Similarly we have a short exact sequence of (multiplicative) abelian groups with $G$-action
\begin{eqnarray}
1 \longrightarrow \mathrm U(1)_T \longrightarrow \mathbb C^\times_T \longrightarrow \mathbb R_+ \longrightarrow 1
\end{eqnarray}
where the subscript $T$ means the complex conjugate $G$-action on coefficient groups. Note that the $G$-action on $\mathbb R_+$ is trivial. This short exact sequence induces a long exact sequence of cohomology groups
\begin{eqnarray}
&& 1 \longrightarrow H^0(G,\mathrm U(1)_T)\longrightarrow H^0(G,\mathbb C^\times_T) \longrightarrow H^0(G,\mathbb R_+)\nonumber \\
&& \phantom{1} \longrightarrow H^1(G,\mathrm U(1)_T)\longrightarrow H^1(G,\mathbb C^\times_T) \longrightarrow H^1(G,\mathbb R_+)\nonumber \\
&& \phantom{1} \longrightarrow H^2(G,\mathrm U(1)_T)\longrightarrow H^2(G,\mathbb C^\times_T) \longrightarrow \cdots\cdots
\end{eqnarray}
Recall that $H^0(G,A_T)$ is the $G$-invariant subgroup of $A$ for all $A$, thus $H^0(G,\mathrm U(1)_T) = \{\pm 1\} = \mathbb Z_2$ and $H^0(G,\mathbb C^\times) = \{z \in \mathbb C^\times \mid z = \bar z\} = \mathbb R^\times$. So three $H^0$ groups in the above sequence form a short exact sequence
\begin{eqnarray}
&& 1 \longrightarrow \mathbb Z_2 \longrightarrow \mathbb R^\times \longrightarrow \mathbb R_+ \longrightarrow 1 .
\end{eqnarray}
Also, all cohomology groups $H^i(G,\mathbb R_+)$ are trivial for $i > 0$. Hence we get short exact sequences
\begin{eqnarray}
1 \longrightarrow H^i(G,\mathrm U(1)_T) \longrightarrow H^i(G,\mathbb C^\times_T) \longrightarrow 1
\end{eqnarray}
for all $i > 0$, which imply that $H^i(G,\mathrm U(1)_T)$ is isomorphic to $H^i(G,\mathbb C^\times_T)$ for all $i > 0$.

A similar statement holds for any compact Lie group $G$, where the Lie group cohomology with coefficients in an abelian Lie group is defined with the differentiable cohomology \cite{brylinski2000differentiable}.  The proof is essentially the same. First, the short exact sequence
$$
0 \longrightarrow \mathrm U(1) \longrightarrow \mathbb C^\times \longrightarrow \mathbb R_+ \longrightarrow 0
$$
of abelian Lie groups induces a long exact sequence of differentiable cohomology groups; then we use the fact $H^i(G,\mathbb R_+) \cong H^i(G,\mathbb R) = 0$ to complete the proof. These two results are listed in Ref. \cite{brylinski2000differentiable}.

Physically, the above results suggest that for generic non-Hermitian systems in 1D, the edge modes of bosonic SPT phase still carry projective representations of the corresponding symmetry group. Unlike Hermitian systems with unitary representation, generic non-Hermitian systems admit non-unitary projective representation. However, the isomorphism $H^2(G;\mathrm U(1)_T) \cong H^2(G;\mathbb C^\times_T)$ suggests that equivalent class of non-unitary projective representations are exactly the same as unitary projective representations. In particular, by a proper gauge choice, e.g., applying a  coboundary transformation, all the projective representations in generic 1D non-Hermitian systems can still be chosen as unitary. Therefore, the classification of bosonic SPT phases in generic 1D non-Hermitian systems is exactly the same as their Hermitian counterparts. Again, as 1D local fermionic systems can always be mapped onto local bosonic systems, the above claim also holds for classifying SPT phases in generic non-Hermitian fermionic systems. Finally, as the above proof shows that $H^i(G;\mathrm U(1)_T) \cong H^i(G;\mathbb C^\times_T)$ for all $i > 0$, we conjecture that in higher dimensions, the classification of bosonic SPT phases in generic non-Hermitian systems is still exactly the same as their Hermitian counterparts.

\subsection{The emergence of unitarity for generic topological phases in 1D non-Hermitian systems} \label{sec:tqft}

Since the low energy effective theory of topological phases can always be described by certain TQFTs, below we discuss a more general way to understand the emergence of unitarity for all topological phases in 1D non-Hermitian systems.
It is well-known that Frobenius algebras give rise to a generic framework to understand 1+1D TQFTs for 1+1D bosonic systems. Based on such a framework, we provide a much deeper understanding on why the classification of topological phases in generic non-Hermitian systems is always the same as their Hermitian counterparts. We show that the concept of emergent unitarity actually applies for all extendable TQFTs in 1+1D.

\subsubsection{Definition of TQFT}

\begin{figure}[htbp]
\[
\begin{array}{c}
\begin{tikzpicture}
\draw (-1,0) .. controls (-1,1) and (0,1) .. (0,2) ;
\draw (2,0) .. controls (2,1) and (1,1) .. (1,2) ;
\draw (0,0) .. controls (0,1) and (1,1) .. (1,0) ;
\fill[gray!10,opacity=0.7] (-1,0) .. controls (-1,1) and (0,1) .. (0,2) .. controls (0,2.3) and (1,2.3) .. (1,2) .. controls (1,1) and (2,1) .. (2,0) .. controls (2,0.3) and (1,0.3) .. (1,0) .. controls (1,1) and (0,1) .. (0,0) .. controls (0,0.3) and (-1,0.3) .. (-1,0) ;
\draw (0,2) .. controls (0,2.3) and (1,2.3) .. (1,2) ;
\draw (-1,0) .. controls (-1,0.3) and (0,0.3) .. (0,0) ;
\draw (1,0) .. controls (1,0.3) and (2,0.3) .. (2,0) ;
\fill[gray!10,opacity=0.7] (-1,0) .. controls (-1,1) and (0,1) .. (0,2) .. controls (0,1.7) and (1,1.7) .. (1,2) .. controls (1,1) and (2,1) .. (2,0) .. controls (2,-0.3) and (1,-0.3) .. (1,0) .. controls (1,1) and (0,1) .. (0,0) .. controls (0,-0.3) and (-1,-0.3) .. (-1,0) ;
\draw (0,2) .. controls (0,1.7) and (1,1.7) .. (1,2) ;
\draw (-1,0) .. controls (-1,-0.3) and (0,-0.3) .. (0,0) ;
\draw (1,0) .. controls (1,-0.3) and (2,-0.3) .. (2,0) ;
\node[left] at (0,2) {$\Sigma_2$} ;
\node[left] at (-0.5,1) {$M$} ;
\node[left] at (-1,0) {$\Sigma_1$} ;
\draw[-stealth,thick] (-2,0.5)--(-2,1.8) node[very near end,right] {Time} ;
\end{tikzpicture}
\end{array}
\rightsquigarrow
\begin{array}{c}
\begin{tikzpicture}
\draw[->] (0,0.2) node[below] {$Z_0(\Sigma_1)$} --(0,1.8) node[above] {$Z_0(\Sigma_2)$} node[midway,left] {$Z_1(M)$} ;
\end{tikzpicture}
\end{array}
\]
\caption{A TQFT maps space and spacetime manifolds to Hilbert spaces and operators.}
\label{fig:TQFT}
\end{figure}
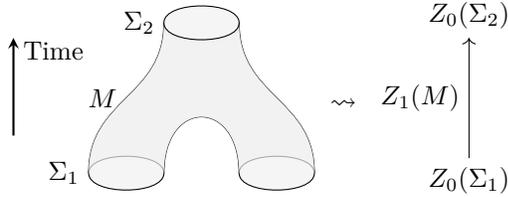

The mathematical definition of a TQFT arises from the intuitions of path integrals. Given a quantum field theory $Z$, there should be a Hilbert space on each (closed) space manifold $\Sigma$, denoted by $Z_0(\Sigma)$; and for each spacetime manifold $M$, whose time slices at the beginning and end are $\Sigma_1$ and $\Sigma_2$ respectively (we call $M$ a cobordism from $\Sigma_1$ to $\Sigma_2$), we can do path integral and get a propagator denoted by $Z_1(M) : Z_0(\Sigma_1) \to Z_0(\Sigma_2)$ (see Fig.~\ref{fig:TQFT}). In particular, if the spacetime manifold $M$ is already closed, $Z_1(M)$ is a complex number, which is just the partition function. Moreover, these data should satisfy some natural factorization properties. For example, if $M$ is a cobordism from $\Sigma_1$ to $\Sigma_2$ and $N$ is a cobordism from $\Sigma_2$ to $\Sigma_3$, then we can sew them together to get a cobordism from $\Sigma_1$ to $\Sigma_3$, denoted by $N \circ M$; then we expect that the composition $Z_1(N) Z_1(M)$ of propagators should be equal to $Z_1(N \circ M)$. We say such a quantum field theory $Z$ is topological if both the Hilbert spaces $Z_0(\Sigma)$ and the propagators $Z_1(M)$ only depend on the topology of the manifolds.
These intuitions lead to a precise mathematical definition of TQFTs\cite{Seg88,Ati88}.

It is well-known that a 1+1D TQFT $Z$ is determined by its value $Z_0(S^1)$ on a circle $S^1$, which is a commutative Frobenius algebra. For example, its multiplication is given by $Z_1(M)$ as depicted in Fig.~\ref{fig:TQFT}. We briefly review the notion of Frobenius algebras in Appendix B.

TQFTs give an abstract but generic framework to understand topological phases. Given a gapped topological phase, its ground state degeneracy defined on a closed space manifold $\Sigma$ is robust and only depends on the topology of $\Sigma$. Therefore, the ground state subspace on $\Sigma$ is well-defined. If its low energy effective theory is described by a TQFT $Z$, then $Z_0(\Sigma)$ is given by this robust ground state subspace. This explains the well-known statement that the low energy effective theories of topological phases are TQFTs. However, a topological phase is an equivalence class of local Hamiltonians, thus only local TQFTs describe topological phases. In the following we discuss the locality of TQFTs.

\subsubsection{Fully extended TQFT}

A TQFT defined as above is not necessarily a local theory. To compute the partition function on an $n$+1D closed manifold, we can decompose it into several pieces along $n$D submanifolds and then do path integrals. However, these pieces are usually too large to be thought of as local. A local theory should provide a way to compute the partition function by any decomposition the spacetime manifold into small enough pieces, for example $n$-simplices, along higher-codimensional submanifolds. Thus a local TQFT should be able to evaluate at not only $n$+1D and $n$D manifolds, but also at all higher-codimensional manifolds. Also, these data should satisfy some compatibility conditions.

This idea leads to the notion of an extended TQFT. Briefly speaking, a fully extended TQFT is an assignment which assigns linear objects (numbers, linear spaces, linear categories, \ldots) to cobordisms of all codimensions, including a point. It is local in the sense that, the partition functions can be computed by the decompositions into arbitrarily small pieces, and, as one may expect, a fully extended TQFT is determined by its value on a point \cite{BD95,Lur08}. If we forget what we assign to higher-codimensional submanifolds, a fully extended TQFT gives an ordinary TQFT. We say a TQFT is fully extendable, or local, if it can be extended to a fully extended TQFT. The classification of fully extended 1+1D TQFTs was given by Ref. \cite{schommerpries2011}, and we summarize the main result in the following:
\begin{thm} \label{thm:ETQFT}
An extended 1+1D TQFT is determined by its value on a point, which is a symmetric separable (semisimple) Frobenius algebra $B$.
\end{thm}
So far we have proved that the low energy effective theory of any local Hamiltonian of a topological phase, no matter it is Hermitian or non-Hermitian, must correspond to a Frobenius algebra $B$.
This Frobenius algebra $B$ encodes all local physical data. In the next subsection we classify the gauged theories of (bosonic) SPT phases with onsite symmetry $G$, and the $G$-gauge symmetries are also encoded in this  algebra $B$.

\medskip
Let us give an explicit lattice construction of a fully extendable 1+1D TQFT from a symmetric special Frobenius algebra $B$ \cite{FHK94}. Fix a basis $\{e_i\}$ of $B$, we write the structure constants of $B$ as:
\begin{align}
\label{structure constants}
e_i \cdot e_j = \mu(e_i \otimes e_j) & = \sum_k C_{ij}^{\phantom{ij}k} \cdot e_k , \\
\Delta(e_i) & = \sum_{jk} C_i^{\phantom{i}jk} \cdot e_k \otimes e_j , \\
\varepsilon(e_i \cdot e_j \cdot e_k) & = C_{ijk} , \\
((\Delta \otimes \id_B) \circ \Delta)(1) & = \sum_{ijk} C^{ijk} \cdot e_k \otimes e_j \otimes e_i .
\end{align}
$\mu$, $\Delta$ and $\varepsilon$ are called multiplication map, comultiplication map and counit map, respectively. See Section B in the Supplementary materials for more details about Frobenius algebra.

For any closed manifold $M$, we take a directed trivalent graph on $M$ and label each edge by basis vectors $\{e_i\}$. Then we assign to each vertex a number as follows:
\begin{align}
F \Biggl(
\begin{array}{c}
\begin{tikzpicture}[scale=0.5]
\draw[->-] (90:1.5)--(0,0) node[midway,right] {$k$} ;
\draw[->-] (-30:1.5)--(0,0) node[midway,below] {$j$} ;
\draw[->-] (210:1.5)--(0,0) node[midway,below] {$i$} ;
\end{tikzpicture}
\end{array} \Biggr) = C_{ijk} , \,
F \Biggl(
\begin{array}{c}
\begin{tikzpicture}[scale=0.5]
\draw[->-] (0,0)--(90:1.5) node[midway,right] {$k$} ;
\draw[->-] (-30:1.5)--(0,0) node[midway,below] {$j$} ;
\draw[->-] (210:1.5)--(0,0) node[midway,below] {$i$} ;
\end{tikzpicture}
\end{array} \Biggr) = C_{ij}^{\phantom{ij}k} , \nonumber \\
F \Biggl(
\begin{array}{c}
\begin{tikzpicture}[scale=0.5]
\draw[->-] (-90:1.5)--(0,0) node[midway,right] {$i$} ;
\draw[->-] (0,0)--(30:1.5) node[midway,above] {$j$} ;
\draw[->-] (0,0)--(150:1.5) node[midway,above] {$k$} ;
\end{tikzpicture}
\end{array} \Biggr) = C_i^{\phantom{i}jk} , \,
F \Biggl(
\begin{array}{c}
\begin{tikzpicture}[scale=0.5]
\draw[->-] (0,0)--(-90:1.5) node[midway,right] {$i$} ;
\draw[->-] (0,0)--(30:1.5) node[midway,above] {$j$} ;
\draw[->-] (0,0)--(150:1.5) node[midway,above] {$k$} ;
\end{tikzpicture}
\end{array} \Biggr) = C^{ijk} .
\end{align}
The partition function on $M$ is obtained by first taking the multiplication of $F(v)$ over all vertices and then summing over all possible labels on basis vectors:
\begin{align}
\Z(M) \coloneqq \sum_F \prod_v F(v) .
\label{FBaction}
\end{align}
This TQFT is local because the lattice construction only involves local data. For more details of this construction, see Section C in the Supplementary materials.

\subsubsection{TQFT with a gauge symmetry}

For any 1+1D SPT phases with onsite(internal) symmetry $G$, we can gauge the symmetry $G$ and get the well known Dijkgraaf-Witten theory, which can be described by a 1+1D TQFT with $G$-gauge symmetry. Thus, there is a one-to-one correspondence between the classification of SPT phases protected by onsite(internal) symmetry $G$  and the classification of Dijkgraaf-Witten theory with $G$-gauge symmetries.

On the other hand, if there is a $G$-gauge symmetry in the partition function \eqref{FBaction}, the local degree of freedom on each edge should be labeled by group elements $g \in G$. In other words, the vector space $B$ is spanned by the group $G$, i.e., $B = \mathbb C[G]$. To do local gauge transformations, the multiplication of $B$ should behave like the group multiplication, up to a phase factor: $g \cdot h = \omega(g,h) \cdot gh$ for some $\omega(g,h) \in \mathbb C^\times$. The associativity of the multiplication implies that $\omega \in Z^2(G;\mathbb C^\times)$ is a 2-cocycle. Hence, the Frobenius algebra $B$ is just the twisted group algebra $\mathbb C_\omega[G]$. Moreover, two cocycles differed by a coboundary give the isomorphic twisted group algebra. Hence for genric non-Hermitian systems, the Dijkgraaf-Witten theory with $G$-gauge symmetry  are classified by the group cohomology $H^2(G;\mathbb C^\times)$, which is isomorphic to $H^2(G;\mathrm U(1))$ by the results in Sec.~\ref{sequence}.  According to the one-to-one correspondence between Dijkgraaf-Witten gauge theory and SPT phases, we end up with the same conclusion that the classification of SPT phases in generic 1+1D non-Hermitian systems is still exactly the same as their Hermitian counterparts. More rigorously, SPT phases in 1+1D can also be classified by $G$-crossed Frobenius algebras and there is a similar cohomology classification \cite{Tur10,MS06}.

Moreover, we note that the Jordan-Wigner transformation maps the Majorana chain (which is a 1D intrinsic topological phase for fermion systems) to a 1d $\mathbb Z_2$ symmetry breaking phase.
It can also be described by the commutative Frobenius algebra $\mathbb C^2 = \mathbb C \oplus \mathbb C$, and the $\mathbb Z_2$-action is permuting two components of $\mathbb C^2$. Apparently, such a commutative Frobenius algebra is always equivalent to a unitary Frobenius algebra with the same $\mathbb Z_2$-action and the concept of emergent unitarity will still apply for this case.

\section{Discussion and conclusion }

We have studied the classification of topological phases for 1D interacting non-Hermitian systems, and established that the classification of topological phases is exactly the same as Hermitian systems. Moreover, unitarity can even emerge for fixed point partition functions of $1$D topological phases. In mathematics, the isomorphisms $H^i(G;\mathrm U(1)_T) \cong H^i(G;\mathbb C^\times_T)$ for all $i > 0$ suggest that in $2$D and $3$D, the classification of interacting SPT phases could still be the same for Hermitian and non-Hermitian systems (at least for bosonic systems). Of course, for intrinsic topological phases in higher dimensions, it has been shown that non-Hermitian systems can be much richer than Hermitian systems, e.g. string-net models constructed by non-unitary fusion category theory are very interesting examples~\cite{lootens2019}.
Finally, defining topological invariants for generic non-Hermitian systems is still quite challenging, which further suggests that topological phase transitions in non-Hermitian systems are much richer than in Hermitian systems, even in $1$D. We believe that non-unitary conformal field theory (CFT) might play a very important role.

\section{Acknowledgments} We are grateful for helpful discussions with Yong-Shi Wu, Meng Cheng, Yang Qi, and Qing-Rui Wang. This work was supported by the National Key Research and Development Program of China (2016YFA0300300), the National Natural Science Foundation of China (NSFC; 11861161001), NSFC/RGC Joint Research Scheme (N-CUHK427/18), the Science, Technology and Innovation Commission of Shenzhen Municipality (ZDSYS20190902092905285), Guangdong Basic and Applied Basic Research Foundation under Grant No. 2020B1515120100 and Center for Computational Science and Engineering of Southern University of Science and Technology.

\appendix

\section{Emergent unitarity for 1+1D intrinsic topological phases in quasi-Hermitian fermionic systems} \label{sec:classification_fermionic}

For $1+1$D quasi-Hermitian fermionic systems, we can use the Grassmann valued amplitude to construct the partition function for topological phases. Below we consider the so-called intrinsic topological phase which is stable even without symmetry protection. It turns out that there is one and only one such kind phase, namely, the Kitaev's Majorana chain model. 

Let us consider the following partition function:

\begin{align}
Z_f=&\sum_{\{n_{ij}\}}\int \prod_{\rm{link}} d \theta^+_{ij}d \theta^-_{ij} \prod_{\rm{link}} (1-\theta^+_{ij} \theta^-_{ij}) \prod_{\rm{triangular}}\mathcal{V}^{s_{ijk}}_{ijk},\label{faction}
\end{align}
where
\begin{align}
\mathcal{V}^+_{ijk}=&\sum_{n_{ij},n_{jk},n_{ik}}
\nu^+(n_{ij},n_{jk},n_{ik})\left(\theta^{+}_{ij}\right)^{n_{ij}}
\left(\theta^{+}_{jk}\right)^{n_{jk}}\left(\theta^{-}_{ik}\right)^{n_{ik}}\nonumber\\
\mathcal{V}^-_{ijk}=&\sum_{n_{ij},n_{jk},n_{ik}}
\nu^-(n_{ij},n_{jk},n_{ik})
\left(\theta^{+}_{ik}\right)^{n_{ik}}\left(\theta^{-}_{jk}\right)^{n_{jk}}\left(\theta^{-}_{ij}\right)^{n_{ij}}
\end{align}
Similar as bosonic systems, for quasi-hermitian systems, we can also impose the following condition: 
\begin{align}
\nu^-(n_{ij},n_{jk},n_{ik})={[\nu^+(n_{ij},n_{jk},n_{ik})]}^*
\end{align}
Note that the fermion parity conservation further requires:
\begin{align}
n_{ij}+n_{jk}+n_{ik}=0 \text{ mod } 2
\end{align}

Actually, in terms of quantum field theory language, 
$\mathcal{V}^{\pm}$ can be regarded as Grassmanm valued amplitude, and the Grassmann
variable $\theta^+/\theta^-$ satisfying standard Grassmann algebra is associate with creation/annihilation operator $c^\dagger/c$.  

Now we consider the time ordered Pachner moves for Grassmann valued partition. Formally, we can write down the $2\leftrightarrow2$ move as:
\begin{widetext}
\begin{align}
\int d \theta^+_{03}d \theta^-_{03} (1-\theta^+_{03} \theta^-_{03})\mathcal{V}^+_{013}\mathcal{V}^-_{023}&=\int d \theta^+_{12}d \theta^-_{12} (1-\theta^+_{12} \theta^-_{12})\mathcal{V}^-_{123}\mathcal{V}^+_{012}
\\
\int d \theta^+_{13}d \theta^-_{13} (1-\theta^+_{13} \theta^-_{13})\mathcal{V}^+_{123}\mathcal{V}^+_{013}&=\int d \theta^+_{02}d \theta^-_{02} (1-\theta^+_{02} \theta^-_{02})\mathcal{V}^+_{023}\mathcal{V}^+_{012}
\end{align}

We note that due to the even number of Grassmann variable constraint, we can remove the summation over $n_{ij}$ and obtain:
\begin{align}
 {\nu^+(n_{01},n_{13},n_{03})}{\nu^-(n_{02},n_{23},n_{03})}=& {\nu^-(n_{12},n_{23},n_{13})}
{\nu^+(n_{01},n_{12},n_{02})}
\\
 {\nu^+(n_{12},n_{23},n_{13})} 
{\nu^+(n_{01},n_{13},n_{03})}=&
 {\nu^+(n_{02},n_{23},n_{03})} 
{\nu^+(n_{01},n_{12},n_{02})} 
\label{1D22move-1}
\end{align}
Similarly, the $1\leftrightarrow3$ moves further imply:
\begin{align}
\nu^+(n_{01},n_{13},n_{03})=&\sum_{n_{12}}
\nu^-(n_{12},n_{23},n_{13}) 
\nu^+(n_{02},n_{23},n_{03})\nu^+(n_{01},n_{12},n_{02})\\
\nu^+(n_{02},n_{23},n_{03})=&\sum_{n_{12}}
\nu^+(n_{12},n_{23},n_{13}) \nu^+(n_{01},n_{13},n_{03}) 
\nu^-(n_{01},n_{12},n_{02}),\label{1D13move}
\end{align}

We notice the combination of the time ordered $2\leftrightarrow 2$
and $1\leftrightarrow3$ moves gives rise to the unitary condition on
$\nu_2^{\pm}$:
\begin{align}
\sum_{n_{ij}}{\nu^{+}(n_{ij},n_{jk},n_{ik})} 
{\nu^{-}(n_{ij},n_{jk},n_{ik}^\prime)}=\delta_{n_{ik},n_{ik}^\prime}
\end{align}
and
\begin{align}
 {\nu^+(n_{12},n_{23},n_{13})} 
{\nu^+(n_{01},n_{13},n_{03})}
 ={\nu^+(n_{02},n_{23},n_{03})} 
{\nu^+(n_{01},n_{12},n_{02})} 
\label{1D22move}
\end{align}
\end{widetext}

A simple solution reads:
\begin{align}
\nu^{\pm}(n_{ij},n_{jk},n_{ik})=1/\sqrt{2}
\label{nosymmetry}
\end{align}
Its corresponding ground state wavefunction(defined by a partion function with a boundary) is
described by an equal weight superposition of all the even number
fermion configurations(associate with a proper fermion ordering).
This solution actually describes the non-trivial phase of Kitaev's Majorana
chain model, which has a protected Majorana zero modes on its open
ends. To see this more explicitly, let us recall the Hamiltonian
of the Kitaev's Majorana chain on an ordered $1$D lattice without a
boundary:
\begin{align}
H=\sum_{I=1}^{N-1}(c_I-c_I^\dagger)(c_{I+1}+c_{I+1}^\dagger)+(c_1-c_1^\dagger)(c_{N}+c_{N}^\dagger)
\end{align}
We choose the anti-periodical
boundary condition(APBC) to simplify our discussion. Due to the fact
${\left[(c_I-c_I^\dagger)(c_{I+1}+c_{I+1}^\dagger)\right]}^2=1$, we
can define the following projectors
\begin{align}
P_I=&\frac{1}{2}[1-(c_I-c_I^\dagger)(c_{I+1}+c_{I+1}^\dagger)];\quad
I=1,\cdots,N-1\nonumber\\
P_N=&\frac{1}{2}[1-(c_1-c_1^\dagger)(c_N+c_N^\dagger)];
\end{align}
It is easy to check $P_I^2=P_I$ and $\left[P_I,P_J\right]=0$. Thus, the Hamiltonian of the Kitaev's Majorana chain model is actually a summation of commuting projectors.
\begin{align}
H=\sum_{I=1}^N \left(1-P_I\right)\label{MajoranaH}
\end{align}
As a result, the ground state can be generated by acting the product
of all projectors $\prod_IP_I$ onto an arbitrary state if the
projector do not annihilate it. It turns out for those states with
even number fermion $\prod_IP_I|\rm{even}\rangle$ do not vanish and
the ground state is an equal weight superposition of all possible
even number fermion configurations(Notice the fermion basis are
ordered as $1<2<\cdots<N$).

In the following, let us explicit show why our solution describes
the fixed point partition function of the Majorana chain. Since the
gap of the system should be infinite at the fixed point, we need to
rescale the Hamiltonian Eq.(\ref{MajoranaH}) as:
\begin{align}
H=U\sum_{I=1}^N \left(1-P_I\right); \quad U \rightarrow \infty
\end{align}
The corresponding fixed point partition function reads:
\begin{align}
Z=e^{-\beta H}=\left(e^{-\Delta\tau H}\right)^n\simeq
\left(\prod_I P_I\right)^n
\end{align}
In the last step we omit the overall constant. In each
imaginary time slice $\Delta\tau=\beta/n$, the partition function takes
a form:
\begin{align}
Z_{\Delta\tau}\simeq \prod_I P_I
\end{align}
In the fermion coherent state representation, we can express each
$P_I$ as:
\begin{align}
\nonumber\\&\langle \theta_I^\prime\theta_{I+1}^\prime|P_I|\theta_I
\theta_{I+1} \rangle\nonumber\\=&\frac{1}{2}\langle
\theta_I^\prime\theta_{I+1}^\prime|\theta_I \theta_{I+1}
\rangle\left[1-(\theta_I-\theta_I^\prime)(\theta_{I+1}+\theta_{I+1}^\prime)\right]\nonumber\\=&\frac{1}{2}
(1+\theta_I^\prime\theta_I)(1+\theta_{I+1}^\prime\theta_{I+1})
\left[1-(\theta_I-\theta_I^\prime)(\theta_{I+1}+\theta_{I+1}^\prime)\right]\nonumber\\=&\frac{1}{2}[
(1+\theta_I^\prime\theta_I)(1+\theta_{I+1}^\prime\theta_{I+1})
-(\theta_I-\theta_I^\prime)(\theta_{I+1}+\theta_{I+1}^\prime)],
\end{align}
where the fermion coherent state $|\theta_I\rangle$ is defined as:
\begin{align}
|\theta_I\rangle=|0\rangle-\theta_Ic_I^\dagger|0\rangle
\end{align}
and from the Grassmann algebra we have $\theta_I^2=0$.

\begin{figure}[h]
\begin{center}
\includegraphics[scale=0.38]{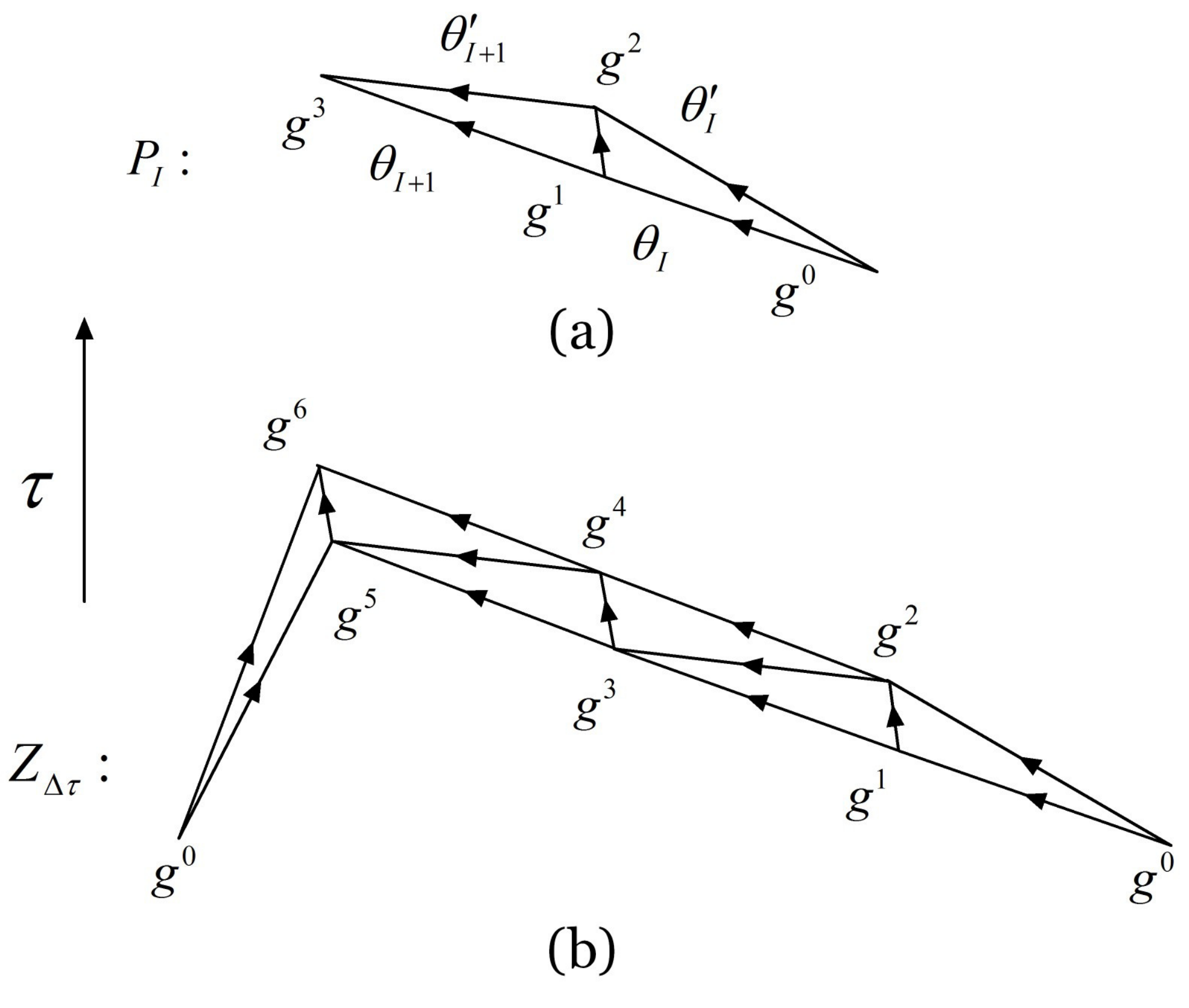}
\end{center}
\caption{The graphic representation of the ideal Hamiltonian for the
Majorana chain. (a) The graphic representation of the projector
$P_I$, all the group elements $g^i$ in this case are just a trivial
identity. (b) The graphic representation for the partition function
$Z_{\Delta\tau}$ for a $3$ sites system with
APBC.\label{1DHamiltonian}}
\end{figure}

The above expression evolve four Grassmann variable, so we need to
introduce two triangle with a shared edge to represent the above
amplitude, see in Fig.~\ref{1DHamiltonian}(a). If we define:
\begin{align}
\theta_I=\theta_{01}^-;\quad \theta_{I+1}=\theta_{13}^-;\quad
\theta_I^\prime=\theta_{02}^+;\quad
\theta_{I+1}^\prime=\theta_{23}^+,
\end{align}
It is easy to check the amplitudes:
\begin{align}
&\int \dd \theta_{12}^+\dd
\theta_{12}^-(1-\theta_{12}^+\theta_{12}^-)
\mathcal{V}^{-}_{012}\mathcal{V}^{+}_{123}\nonumber\\=&
\frac{1}{2}\left[(1+\theta_I^\prime\theta_I)(1+\theta_{I+1}^\prime\theta_{I+1})
-(\theta_I-\theta_I^\prime)(\theta_{I+1}+\theta_{I+1}^\prime)\right]
\nonumber\\ =&\langle
\theta_I^\prime\theta_{I+1}^\prime|P_I|\theta_I \theta_{I+1}
\rangle,
\end{align}
where the coefficients of $\mathcal{V}^{\pm}$ are the
solutions in Eq.(\ref{nosymmetry}):
\begin{align}
\mathcal{V}^{+}_{ijk}=&\frac{1}{\sqrt{2}}\sum_{n_{ij},n_{jk}}
\left(\theta^{+}_{ij}\right)^{n_{ij}}
\left(\theta^{+}_{jk}\right)^{n_{jk}}\left(\theta^{-}_{ik}\right)^{|n_{ij}-n_{jk}|}\nonumber\\
\mathcal{V}^{-}_{ijk}=&\frac{1}{\sqrt{2}}\sum_{n_{ij},n_{jk}}
\left(\theta^{+}_{ik}\right)^{|n_{ij}-n_{jk}|}\left(\theta^{-}_{jk}\right)^{n_{jk}}\left(\theta^{-}_{ij}\right)^{n_{ij}}
\end{align}

Thus, we prove that the amplitude in Fig.~\ref{1DHamiltonian}(a) does
represent the projector $P_I$. The partition function in a time
slice can be constructed by a product of $P_I$, as shown in Fig.~\ref{1DHamiltonian}(b), notice that in a partition function with a global time
ordered structure will be naturally associated with APBC, as discussed above. Other boundary conditions require the introducing of discrete spin structures, which is much more complicated and beyond the scope of this paper. In conclusion, we see that without any physical symmetry, there is still
a non-trivial topological phase in $1$D quasi-Hermitian fermion systems, and unitarity will emerge for its fixed point partition function, which exactly describes the ground state phase of Kitaev's Majorana chain model.

\section{Frobenius algebras} \label{sec:Frobenius}

In this appendix, we briefly review the definition and basic properties of Frobenius algebras.

\begin{defn}
A Frobenius algebra is a quintuple $(A,\mu,\eta,\Delta,\varepsilon)$, where:
\begin{enumerate}
    \item $A$ is a (complex) vector space.
    \item $\mu : A \otimes A \to A$ is a linear map, called the multiplication map. We also denote $\mu(a \otimes b)$ by $a \cdot b$ or simply $ab$.
    \item $\eta : \mathbb C \to A $ is a linear map, called the unit map. We also denote $\eta(1)$ by $1_A$ or simply $1$.
    \item $\Delta : A \to A \otimes A$ is a linear map, called the comultiplication map.
    \item $\varepsilon : A \to \mathbb C$ is a linear map, called the counit map.
\end{enumerate}
These data satisfy the following conditions:
\begin{itemize}
    \item $(A,\mu,\eta)$ is an algebra. More precisely, we have
    \begin{align*}
        \mu \circ (\mu \otimes \id_A) = \mu \circ (\id_A \otimes \mu) , \\
        \mu \circ (\eta \otimes \id_A) = \id_A = \mu \circ (\id_A \otimes \eta) .
    \end{align*}
    Here $\id_A : A \to A$ is the identity map.
    \item $(A,\Delta,\varepsilon)$ is a coalgebra. More precisely, we have
    \begin{align*}
        (\Delta \otimes \id_A) \circ \Delta = (\id_A \otimes \Delta) \circ \Delta , \\
        (\varepsilon \otimes \id_A) \circ \Delta = \id_A = (\id_A \otimes \varepsilon) \circ \Delta .
    \end{align*}
    \item We have
    \[
    (\mu \otimes \id_A) \circ (\id_A \otimes \Delta) = \Delta \circ \mu = (\id_A \otimes \mu) \circ (\Delta \otimes \id_A) .
    \]
    This condition is called the Frobenius condition.
\end{itemize}
\end{defn}

\begin{rem}
By applying the first condition of that $(A,\mu,\eta)$ is an algebra to an element $a \otimes b \otimes c \in A \otimes A \otimes A$, we get $(ab)c = a(bc)$. In other words, this condition means the multiplication is associative. Similarly, the second condition says $1 \cdot a = a = a \cdot 1$ for all $a \in A$, which means $1 \in A$ is the unit for the multiplication. These two conditions reformulate the classical definitions of (linear) algebras by using linear maps, not elements in vector spaces.
\end{rem}

Let us explain the definition of Frobenius algebras by the so-called graph calculus. First, one may imagine that $A$ is a particle moving in the spacetime. So we can draw the Feynman diagrams of $A$. If there is only a single $A$, we draw it as follows:
\[
A =
\begin{array}{c}
\begin{tikzpicture}[scale=0.5]
\draw (0,-1)--(0,1) ;
\end{tikzpicture}
\end{array}
= \id_A .
\]
Our convention is that the time axis points upward. This single line of $A$ also represents the identity map $\id_A$ because $\id_A$ means nothing happens on $A$. All the other data in the definition of Frobenius algebras can be viewed as interactions of $A$, depicted as follows:
\[
\mu =
\begin{array}{c}
\begin{tikzpicture}[scale=0.5]
\draw (-0.8,-1) .. controls (-0.8,-0.5) and (-0.4,0) .. (0,0) ;
\draw (0.8,-1) .. controls (0.8,-0.5) and (0.4,0) .. (0,0) ;
\draw (0,0)--(0,1) ;
\end{tikzpicture}
\end{array} , \quad
\eta =
\begin{array}{c}
\begin{tikzpicture}[scale=0.5]
\draw (0,0)--(0,1) ;
\draw[fill=white] (0,0) circle (0.1) ;
\end{tikzpicture}
\end{array} , \quad
\Delta =
\begin{array}{c}
\begin{tikzpicture}[scale=0.5]
\draw (0,-1)--(0,0) ;
\draw (0,0) .. controls (-0.4,0) and (-0.8,0.5) .. (-0.8,1) ;
\draw (0,0) .. controls (0.4,0) and (0.8,0.5) .. (0.8,1) ;
\end{tikzpicture}
\end{array} , \quad
\varepsilon =
\begin{array}{c}
\begin{tikzpicture}[scale=0.5]
\draw (0,0)--(0,1) ;
\draw[fill=white] (0,1) circle (0.1) ;
\end{tikzpicture}
\end{array} .
\]
Here we use a small circle to depict the one-dimensional vector space $\mathbb C$, which represents the vacuum. Then the conditions of that $(A,\mu,\eta)$ is an algebra can be depicted as
\[
\begin{array}{c}
\begin{tikzpicture}[scale=0.5]
\draw (-1.6,0) .. controls (-1.6,0.5) and (-1.2,1) .. (-0.8,1) ;
\draw (0,0) .. controls (0,0.5) and (-0.4,1) .. (-0.8,1) ;
\draw (-0.8,1) .. controls (-0.8,1.5) and (-0.4,2) .. (0,2) ;
\draw (0.8,0)--(0.8,1) .. controls (0.8,1.5) and (0.4,2) .. (0,2) ;
\draw (0,2)--(0,3) ;
\end{tikzpicture}
\end{array}
=
\begin{array}{c}
\begin{tikzpicture}[scale=0.5]
\draw (-0.8,0)--(-0.8,1) .. controls (-0.8,1.5) and (-0.4,2) .. (0,2) ;
\draw (0,0) .. controls (0,0.5) and (0.4,1) .. (0.8,1) ;
\draw (1.6,0) .. controls (1.6,0.5) and (1.2,1) .. (0.8,1) ;
\draw (0.8,1) .. controls (0.8,1.5) and (0.4,2) .. (0,2) ;
\draw (0,2)--(0,3) ;
\end{tikzpicture}
\end{array} , \quad
\begin{array}{c}
\begin{tikzpicture}[scale=0.5]
\draw (-0.8,-1) .. controls (-0.8,-0.5) and (-0.4,0) .. (0,0) ;
\draw (0.8,-1) .. controls (0.8,-0.5) and (0.4,0) .. (0,0) ;
\draw (0,0)--(0,1) ;
\draw[fill=white] (-0.8,-1) circle (0.1) ;
\end{tikzpicture}
\end{array}
=
\begin{array}{c}
\begin{tikzpicture}[scale=0.5]
\draw (0,-1)--(0,1) ;
\end{tikzpicture}
\end{array}
=
\begin{array}{c}
\begin{tikzpicture}[scale=0.5]
\draw (-0.8,-1) .. controls (-0.8,-0.5) and (-0.4,0) .. (0,0) ;
\draw (0.8,-1) .. controls (0.8,-0.5) and (0.4,0) .. (0,0) ;
\draw (0,0)--(0,1) ;
\draw[fill=white] (0.8,-1) circle (0.1) ;
\end{tikzpicture}
\end{array} ,
\]
i.e. $\mu$ is associative and $\eta$ is unital for $\mu$; the conditions of that $(A,\Delta,\varepsilon)$ is a coalgebra can be depicted as
\[
\begin{array}{c}
\begin{tikzpicture}[scale=0.5]
\draw (0,0)--(0,1) ;
\draw (0,1) .. controls (-0.4,1) and (-0.8,1.5) .. (-0.8,2) ;
\draw (-0.8,2) .. controls (-1.2,2) and (-1.6,2.5) .. (-1.6,3) ;
\draw (-0.8,2) .. controls (-0.4,2) and (0,2.5) .. (0,3) ;
\draw (0,1) .. controls (0.4,1) and (0.8,1.5) .. (0.8,2)--(0.8,3) ;
\end{tikzpicture}
\end{array}
=
\begin{array}{c}
\begin{tikzpicture}[scale=0.5]
\draw (0,0)--(0,1) ;
\draw (0,1) .. controls (-0.4,1) and (-0.8,1.5) .. (-0.8,2)--(-0.8,3) ;
\draw (0,1) .. controls (0.4,1) and (0.8,1.5) .. (0.8,2) ;
\draw (0.8,2) .. controls (0.4,2) and (0,2.5) .. (0,3) ;
\draw (0.8,2) .. controls (1.2,2) and (1.6,2.5) .. (1.6,3) ;
\end{tikzpicture}
\end{array} , \quad
\begin{array}{c}
\begin{tikzpicture}[scale=0.5]
\draw (0,-1)--(0,0) ;
\draw (0,0) .. controls (-0.4,0) and (-0.8,0.5) .. (-0.8,1) ;
\draw (0,0) .. controls (0.4,0) and (0.8,0.5) .. (0.8,1) ;
\draw[fill=white] (-0.8,1) circle (0.1) ;
\end{tikzpicture}
\end{array}
=
\begin{array}{c}
\begin{tikzpicture}[scale=0.5]
\draw (0,-1)--(0,1) ;
\end{tikzpicture}
\end{array}
=
\begin{array}{c}
\begin{tikzpicture}[scale=0.5]
\draw (0,-1)--(0,0) ;
\draw (0,0) .. controls (-0.4,0) and (-0.8,0.5) .. (-0.8,1) ;
\draw (0,0) .. controls (0.4,0) and (0.8,0.5) .. (0.8,1) ;
\draw[fill=white] (0.8,1) circle (0.1) ;
\end{tikzpicture}
\end{array} ,
\]
i.e. $\Delta$ is coassociative and $\varepsilon$ is counital for $\Delta$; and the Frobenius condition can be depicted as
\[
\begin{array}{c}
\begin{tikzpicture}[scale=0.5]
\draw (-1.6,0)--(-1.6,2) .. controls (-1.6,2.5) and (-1.2,3) .. (-0.8,3) ;
\draw (0.8,0)--(0.8,1) ;
\draw (0.8,1) .. controls (0.4,1) and (0,1.5) .. (0,2) .. controls (0,2.5) and (-0.4,3) .. (-0.8,3) ;
\draw (-0.8,3)--(-0.8,4) ;
\draw (0.8,1) .. controls (1.2,1) and (1.6,1.5) .. (1.6,2)--(1.6,4) ;
\end{tikzpicture}
\end{array}
=
\begin{array}{c}
\begin{tikzpicture}[scale=0.5]
\draw (-0.8,0) .. controls (-0.8,0.5) and (-0.4,1) .. (0,1) ;
\draw (0.8,0) .. controls (0.8,0.5) and (0.4,1) .. (0,1) ;
\draw (0,1)--(0,3) ;
\draw (0,3) .. controls (-0.4,3) and (-0.8,3.5) .. (-0.8,4) ;
\draw (0,3) .. controls (0.4,3) and (0.8,3.5) .. (0.8,4) ;
\end{tikzpicture}
\end{array}
=
\begin{array}{c}
\begin{tikzpicture}[scale=0.5]
\draw (-0.8,0)--(-0.8,1) ;
\draw (-0.8,1) .. controls (-1.2,1) and (-1.6,1.5) .. (-1.6,2)--(-1.6,4) ;
\draw (-0.8,1) .. controls (-0.4,1) and (0,1.5) .. (0,2) .. controls (0,2.5) and (0.4,3) .. (0.8,3) ;
\draw (1.6,0)--(1.6,2) .. controls (1.6,2.5) and (1.2,3) .. (0.8,3) ;
\draw (0.8,3)--(0.8,4) ;
\end{tikzpicture}
\end{array} .
\]
Hence the definition of Frobenius algebras means the value of topologically equivalent diagrams are equal. This is also why 1+1D TQFTs are equivalent to commutative Frobenius algebras.

\begin{defn}
A Frobenius algebra $A = (A,\mu,\eta,\Delta,\varepsilon)$ is commutative if it satisfies the condition $\mu \circ \tau = \mu$, where $\tau : A \otimes A \to A \otimes A$ permutes two components: $\tau(a \otimes b) = b \otimes a$. It is symmetric if it satisfies the conditio $\varepsilon \circ \mu \circ \tau = \varepsilon \circ \mu$.
\end{defn}

We can also depict the map $\tau$ as
\[
\tau =
\begin{array}{c}
\begin{tikzpicture}[scale=0.5]
\draw (-0.8,-1) .. controls (-0.8,0) and (0.8,0) .. (0.8,1) ;
\draw (0.8,-1) .. controls (0.8,0) and (-0.8,0) .. (-0.8,1) ;
\end{tikzpicture}
\end{array} .
\]
Then $A$ is commutative can be depicted as
\[
\begin{array}{c}
\begin{tikzpicture}[scale=0.5]
\draw (0.8,-3) .. controls (0.8,-2) and (-0.8,-2) .. (-0.8,-1) .. controls (-0.8,-0.5) and (-0.4,0) .. (0,0) ;
\draw (-0.8,-3) .. controls (-0.8,-2) and (0.8,-2) .. (0.8,-1) .. controls (0.8,-0.5) and (0.4,0) .. (0,0) ;
\draw (0,0)--(0,1) ;
\end{tikzpicture}
\end{array}
=
\begin{array}{c}
\begin{tikzpicture}[scale=0.5]
\draw (-0.8,-3)--(-0.8,-1) .. controls (-0.8,-0.5) and (-0.4,0) .. (0,0) ;
\draw (0.8,-3)--(0.8,-1) .. controls (0.8,-0.5) and (0.4,0) .. (0,0) ;
\draw (0,0)--(0,1) ;
\end{tikzpicture}
\end{array} ,
\]
and that $A$ is symmetric can be depicted as
\[
\begin{array}{c}
\begin{tikzpicture}[scale=0.5]
\draw (0.8,-3) .. controls (0.8,-2) and (-0.8,-2) .. (-0.8,-1) .. controls (-0.8,-0.5) and (-0.4,0) .. (0,0) ;
\draw (-0.8,-3) .. controls (-0.8,-2) and (0.8,-2) .. (0.8,-1) .. controls (0.8,-0.5) and (0.4,0) .. (0,0) ;
\draw (0,0)--(0,1) ;
\draw[fill=white] (0,1) circle (0.1) ;
\end{tikzpicture}
\end{array}
=
\begin{array}{c}
\begin{tikzpicture}[scale=0.5]
\draw (-0.8,-3)--(-0.8,-1) .. controls (-0.8,-0.5) and (-0.4,0) .. (0,0) ;
\draw (0.8,-3)--(0.8,-1) .. controls (0.8,-0.5) and (0.4,0) .. (0,0) ;
\draw (0,0)--(0,1) ;
\draw[fill=white] (0,1) circle (0.1) ;
\end{tikzpicture}
\end{array} .
\]

%

\begin{expl}
Let us consider the case that the commutative Frobenius algebra $A$ is semisimple. So as an algebra we have $A \simeq \mathbb C^n$ for some positive integer $n$, and the Frobenius structure over $\mathbb C^n$ is determined by the counit map $\varepsilon : \mathbb C^n \to \mathbb C$. Let $\{e_i = (0,\ldots,1,\ldots,0) \in \mathbb C^n\}$ be the canonical basis of $\mathbb C^n$, and denote $\theta_i = \varepsilon(e_i)$. They are $n$ non-zero numbers. Then the coproduct $\Delta : \mathbb C^n \to \mathbb C^n \otimes \mathbb C^n$ is given by $\Delta(e_i) = \theta_i^{-1} \cdot e_i \otimes e_i$.

This commutative Frobenius algebra $A$ corresponds to a 1+1D TQFT, so we can compute the partition functions of the corresponding TQFT on closed surfaces. The results are listed below.
\begin{itemize}
    \item The partition function on a sphere is equal to $\varepsilon \circ \eta = \varepsilon(1) = \sum_{i=1}^n \theta_i$.
    \item The partition function on a torus is equal to $\varepsilon \circ \mu \circ \Delta \circ \eta = \varepsilon \circ \mu (\Delta(1)) = \varepsilon \circ \mu (\sum_{i=1}^n \theta_i^{-1} \cdot e_i \otimes e_i) = \varepsilon(\sum_{i=1}^n \theta_i^{-1} \cdot e_i) = \sum_{i=1}^n 1 = n$, which is the dimension of $A$.
    \item The partition function on a genus-$g$ surface is equal to $\sum_{i=1}^n \theta_i^{1-g}$.
\end{itemize}
\end{expl}

\section{State-sum constructions of 1+1D TQFTs} \label{sec:state-sum}

In this appendix we review a state-sum construction of 1+1D TQFTs based on Frobenius algebras\cite{FHK94}. For simplicity, we only give the construction of partition functions on closed surfaces $M$.
First we fix a symmetric special Frobenius algebra $B = (B,\mu,\eta,\Delta,\varepsilon)$. We also choose a basis $\{e_i\}$ of $B$, then all linear maps $\mu,\eta,\Delta,\varepsilon$ can be equivalently represented by structure constants in this basis. For example, we define $\{C_{ij}^{\phantom{ij}k}\}$ by
\be
e_i \cdot e_j = \mu(e_i \otimes e_j) = \sum_k C_{ij}^{\phantom{ij}k} \cdot e_k ,
\ee
and $\{C_i^{\phantom{i}jk}\}$ by
\be
\Delta(e_i) = \sum_{jk} C_i^{\phantom{i}jk} \cdot e_k \otimes e_j .
\ee
We also define $\{C_{ijk}\}$ by
\be
\varepsilon(e_i \cdot e_j \cdot e_k) = C_{ijk} ,
\ee
and $\{C^{ijk}\}$ by
\be
((\Delta \otimes \id_B) \circ \Delta)(1) = \sum_{ijk} C^{ijk} \cdot e_k \otimes e_j \otimes e_i .
\ee
These coefficients are used to construct partition functions.

Given a 2-dimensional compact oriented manifold $M$ without boundary, we choose a trivalent graph $K'$ on $M$. This can be done by first taking a triangulation of $M$ and then taking the dual graph. We also need to order the graph, but in this case the order can be weaker than a branching structure: what we need is only an order of each edge, that is, a directed graph.

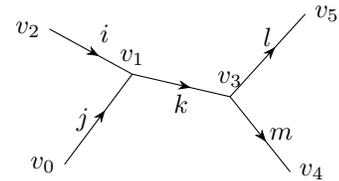
\begin{figure}[htbp]
\centering
\begin{tikzpicture}
\coordinate[label=left:$v_0$] (v0) at (0,0) ;
\coordinate[label=above:$v_1$] (v1) at (0.9,1.2) ;
\coordinate[label=left:$v_2$] (v2) at (-0.2,1.8) ;
\coordinate[label=above:$v_3$] (v3) at (2.2,0.9) ;
\coordinate[label=right:$v_4$] (v4) at (3,-0.1) ;
\coordinate[label=right:$v_5$] (v5) at (3.2,2) ;
\draw[->-] (v0)--(v1) node[midway,left] {$j$} ;
\draw[->-] (v2)--(v1) node[midway,above right] {$i$} ;
\draw[->-] (v1)--(v3) node[midway,below] {$k$} ;
\draw[->-] (v3)--(v4) node[midway,right] {$m$} ;
\draw[->-] (v3)--(v5) node[midway,above] {$l$} ;
\end{tikzpicture}
\caption{a directed trivalent graph with labels}
\label{fig:directed_graph}
\end{figure}

A $B$-field $F$ is an assignment to each oriented edge $(v_\alpha,v_\beta)$ a basis vector $F(v_\alpha,v_\beta) \in \{e_i\}$ of $B$ (see Figure~\ref{fig:directed_graph}). Given a $B$-field, we assign each vertex $v$ a number $F(v)$ as follows:

\begin{align*}
F \Biggl(
\begin{array}{c}
\begin{tikzpicture}[scale=0.5]
\draw[->-] (90:1.5)--(0,0) node[midway,right] {$k$} ;
\draw[->-] (-30:1.5)--(0,0) node[midway,below] {$j$} ;
\draw[->-] (210:1.5)--(0,0) node[midway,below] {$i$} ;
\end{tikzpicture}
\end{array} \Biggr) = C_{ijk} , \,
F \Biggl(
\begin{array}{c}
\begin{tikzpicture}[scale=0.5]
\draw[->-] (0,0)--(90:1.5) node[midway,right] {$k$} ;
\draw[->-] (-30:1.5)--(0,0) node[midway,below] {$j$} ;
\draw[->-] (210:1.5)--(0,0) node[midway,below] {$i$} ;
\end{tikzpicture}
\end{array} \Biggr) = C_{ij}^{\phantom{ij}k} , \\
F \Biggl(
\begin{array}{c}
\begin{tikzpicture}[scale=0.5]
\draw[->-] (-90:1.5)--(0,0) node[midway,right] {$i$} ;
\draw[->-] (0,0)--(30:1.5) node[midway,above] {$j$} ;
\draw[->-] (0,0)--(150:1.5) node[midway,above] {$k$} ;
\end{tikzpicture}
\end{array} \Biggr) = C_i^{\phantom{i}jk} , \,
F \Biggl(
\begin{array}{c}
\begin{tikzpicture}[scale=0.5]
\draw[->-] (0,0)--(-90:1.5) node[midway,right] {$i$} ;
\draw[->-] (0,0)--(30:1.5) node[midway,above] {$j$} ;
\draw[->-] (0,0)--(150:1.5) node[midway,above] {$k$} ;
\end{tikzpicture}
\end{array} \Biggr) = C^{ijk} .
\end{align*}
Note that the order of indices depends on the orientation of the manifold $M$, and in the above figures we assume the orientation is counterclockwise. Since $B$ is symmetric, these structure constants are cyclic invariant, for example, $C_{ijk} = C_{jki} = C_{kij}$, thus these numbers $F(v)$ are well-defined. Then we define a number $Z(K',F,B)$ to be the multiplication of all numbers assigned to vertices:
\[
Z(K',F,B) \coloneqq \prod_v F(v) .
\]

The partition function on $M$ is obtained by summing over all $B$-fields, i.e. summing over all indices of basis vectors:
\be \label{eq:partition_function}
\Z = \Z(K',B) \coloneqq \sum_F Z(K',F,B) = \sum_F \prod_v F(v) .
\ee
For example, the $B$-field in Figure~\ref{fig:directed_graph} gives $Z = C_{ij}^{\phantom{ij}k} C_k^{\phantom{k}lm} \cdots$, and the partition function should be
\[
\Z = \sum_{ijklm\cdots} C_{ij}^{\phantom{ij}k} C_k^{\phantom{k}lm} \cdots .
\]
Since all indices are summing over, the partition function is independent of the choice of the basis $\{e_i\}$ of $B$.

\medskip
Before we prove the topological invariance of the partition function \eqref{eq:partition_function}, we give some examples which recover the Dijkgraaf-Witten theory.

\begin{expl}
Let $G$ be a finite group. Define a Frobenius algebra $B = \mathbb C[G]$ as follow:
\begin{enumerate}
    \item The vector space $\mathbb C[G]$ is spanned by the elements in $G$. In other words, the vectors in $\mathbb C[G]$ are of the form $\sum_{g \in G} \lambda_g \cdot g$.
    \item The multiplication of $\mathbb C[G]$ is given by the group multiplication of $G$: $\mu(g \otimes h) = gh$.
    \item The unit of $\mathbb C[G]$ is the unit of $G$: $\eta(1) = e$.
    \item The comultiplication of $\mathbb C[G]$ is defined by $\Delta(g) = \vert G \vert^{-1} \sum_{xy = g} x \otimes y$.
    \item The counit of $\mathbb C[G]$ is defined by $\varepsilon(g) = \delta_{g,e} \cdot \vert G \vert$.
\end{enumerate}
It is not hard to verify that $\mathbb C[G]$ is a symmetric special Frobenius algebra. We choose the canonical basis $\{g \in \mathbb C[G]\}_{g \in G}$, then we have
\begin{align*}
C_{gh}^{\phantom{gh}k} = \delta_{gh}^k , \, C_g^{\phantom{g}hk} = \lvert G \rvert^{-1} \delta_g^{kh} , \\
C_{ghk} = \lvert G \rvert \delta_{ghk}^e , \, C^{ijk} = \lvert G \rvert^{-2} \delta^{khg}_e ,
\end{align*}
where $\delta_x^y$ is the delta function. Given a $\mathbb C[G]$-field $F$, it is clear that $Z(K',F,\mathbb C[G])$ is nonzero only if for each vertices the multiplication of three adjacent group elements (up to orientation) is $e \in G$. On the dual triangulation this condition is nothing but the flat-connection condition. It is not hard to verify that the partition function coincides with the Dijkgraaf-Witten theory with the trivial cocycle $\omega = 1$.
\end{expl}

\begin{expl}
The above example can be generalized to a twisted version. Suppose $\omega \in Z^2(G;\mathbb C^\times)$ is a $2$-cocycle of the group cohomology valued in $\mathbb C^\times$. Define a Frobenius algebra $B = \mathbb C_\omega[G]$ as follow:
\begin{enumerate}
    \item The vector space $\mathbb C_\omega[G]$ is the same as $\mathbb C[G]$.
    \item The multiplication of $\mathbb C_\omega[G]$ is given by $\mu(g \otimes h) = \omega(g,h) \cdot gh$.
    \item The unit of $\mathbb C_\omega[G]$ is $\eta(1) = \omega(e,e)^{-1} \cdot e$. Recall that the cocycle condition implies that $\omega(g,e) = \omega(e,e) = \omega(e,g)$ for all $g \in G$. Usually we take $\omega$ to be normalized, i.e. $\omega(e,e) = 1$.
    \item The comultiplication of $\mathbb C_\omega[G]$ is defined by $\Delta(g) = \vert G \vert^{-1} \sum_{xy = g} \omega(x,y)^{-1} \cdot x \otimes y$.
    \item The counit of $\mathbb C_\omega[G]$ is defined by $\varepsilon(g) = \delta_{g,e} \cdot \vert G \vert \cdot \omega(e,e)$.
\end{enumerate}
As the above example, it can be verified that the partition function $\Z(K',\mathbb C_\omega[G])$ coincides with the Dijkgraaf-Witten theory.
\end{expl}

\begin{rem}
The partition function \eqref{eq:partition_function} can be written in a basis-free version. We can view each edge as a copy of $B$ and each vertex as a linear map. For example, the vertex
\[
\begin{tikzpicture}[scale=0.5]
\draw[->-] (0,0)--(90:1.5) ;
\draw[->-] (-30:1.5)--(0,0) ;
\draw[->-] (210:1.5)--(0,0) ;
\end{tikzpicture}
\]
should be viewed as the linear map $\mu : B \otimes B \to B$. Then the whole graph should be viewed as a trace of a large linear map, or equivalently a linear map from $\mathbb C$ to $\mathbb C$, since there is no outer legs. This trace is precisely the partition function \eqref{eq:partition_function}.
\end{rem}

\begin{rem}
It is not necessary to use a trivalent graph to do this construction. From a Frobenius algebra $B$ we can easily construct a linear map $B^{\otimes m} \to B^{\otimes n}$ for any $m,n$. Hence this construction works for a graph whose vertices has arbitrarily many edges.
\end{rem}

\medskip
Let us prove the topological invariance of the partition function \eqref{eq:partition_function}.

\medskip
It suffices to show that the partition function is invariant under (dual) Pachner moves.
The $(2,2)$-moves are
\begin{align*}
\begin{array}{c}
\begin{tikzpicture}[scale=0.5]
\draw[->-] (0,0)--(90:1.5) ;
\draw[->-] (-30:1.5)--(0,0) ;
\draw[->-] (210:1.5)--(0,0) ;
\draw[->-] (90:1.5)+(30:1.5)--(90:1.5) ;
\draw[->-] (90:1.5)--+(150:1.5) ;
\end{tikzpicture}
\end{array}
& \leftrightsquigarrow
\begin{array}{c}
\begin{tikzpicture}[scale=0.5]
\draw[->-] (0,0)--(120:1.5) ;
\draw[->-] (240:1.5)--(0,0) ;
\draw[->-] (0:1.5)--(0,0) ;
\draw[->-] (0:1.5)+(60:1.5)--(0:1.5) ;
\draw[->-] (0:1.5)+(-60:1.5)--(0:1.5) ;
\end{tikzpicture}
\end{array}
\text{ or }
\begin{array}{c}
\begin{tikzpicture}[scale=0.5]
\draw[->-] (0,0)--(120:1.5) ;
\draw[->-] (240:1.5)--(0,0) ;
\draw[->-] (0,0)--(0:1.5) ;
\draw[->-] (0:1.5)+(60:1.5)--(0:1.5) ;
\draw[->-] (0:1.5)+(-60:1.5)--(0:1.5) ;
\end{tikzpicture}
\end{array} , \\
\begin{array}{c}
\begin{tikzpicture}[scale=0.5]
\draw[->-] (0,0)--(90:1.5) ;
\draw[->-] (-30:1.5)--(0,0) ;
\draw[->-] (210:1.5)--(0,0) ;
\draw[->-] (90:1.5)--+(30:1.5) ;
\draw[->-] (90:1.5)--+(150:1.5) ;
\end{tikzpicture}
\end{array}
& \leftrightsquigarrow
\begin{array}{c}
\begin{tikzpicture}[scale=0.5]
\draw[->-] (0,0)--(120:1.5) ;
\draw[->-] (240:1.5)--(0,0) ;
\draw[->-] (0,0)--(0:1.5) ;
\draw[->-] (0:1.5)--+(60:1.5) ;
\draw[->-] (0:1.5)+(-60:1.5)--(0:1.5) ;
\end{tikzpicture}
\end{array}
\text{ or }
\begin{array}{c}
\begin{tikzpicture}[scale=0.5]
\draw[->-] (0,0)--(120:1.5) ;
\draw[->-] (240:1.5)--(0,0) ;
\draw[->-] (0:1.5)--(0,0) ;
\draw[->-] (0:1.5)--+(60:1.5) ;
\draw[->-] (0:1.5)+(-60:1.5)--(0:1.5) ;
\end{tikzpicture}
\end{array} , \ldots ,
\end{align*}
i.e. the usual $(2,2)$-move with all possible orientations on edges. Similarly, the $(1,3)$-moves are
\[
\begin{array}{c}
\begin{tikzpicture}[scale=0.5]
\draw[->-] (0,0)--(90:1.5) ;
\draw[->-] (-30:1.5)--(0,0) ;
\draw[->-] (210:1.5)--(0,0) ;
\end{tikzpicture}
\end{array}
\leftrightsquigarrow
\begin{array}{c}
\begin{tikzpicture}[scale=0.5]
\draw[->-] (0,0)--(0:1.5) ;
\draw[->-] (0,0)--(60:1.5) ;
\draw[->-] (210:1.5)--(0,0) ;
\draw[->-] (0:1.5)--+(120:1.5) ;
\draw[->-] (0:1.5)++(-30:1.5)--(0:1.5) ;
\draw[->-] (60:1.5)--+(90:1.5) ;
\end{tikzpicture}
\end{array}
\text{ or }
\begin{array}{c}
\begin{tikzpicture}[scale=0.5]
\draw[->-] (0:1.5)--(0,0) ;
\draw[->-] (0,0)--(60:1.5) ;
\draw[->-] (210:1.5)--(0,0) ;
\draw[->-] (0:1.5)--+(120:1.5) ;
\draw[->-] (0:1.5)++(-30:1.5)--(0:1.5) ;
\draw[->-] (60:1.5)--+(90:1.5) ;
\end{tikzpicture}
\end{array}
\text{ or } \cdots .
\]

The invariance under the first $(2,2)$-move is precisely the associativity of the multiplication $\mu$. Using structure constants, this condition can be written as
\[
C_{ij}^{\phantom{ij}l} C_{lk}^{\phantom{lk}m} = C_{il}^{\phantom{il}m} C_{jk}^{\phantom{jk}l} .
\]
The second $(2,2)$-move is precisely the Frobenius condition. Similarly, all $(2,2)$-moves are related to the associativity, the coassociativity and the Frobenius condition.

For the $(1,3)$-moves, first we apply a $(2,2)$-move to the right hand side of $(1,3)$-moves:
\[
\begin{array}{c}
\begin{tikzpicture}[scale=0.5]
\draw[->-] (0,0)--(0:1.5) ;
\draw[->-] (0,0)--(60:1.5) ;
\draw[->-] (210:1.5)--(0,0) ;
\draw[->-] (0:1.5)--+(120:1.5) ;
\draw[->-] (0:1.5)++(-30:1.5)--(0:1.5) ;
\draw[->-] (60:1.5)--+(90:1.5) ;
\end{tikzpicture}
\end{array}
\leftrightsquigarrow
\begin{array}{c}
\begin{tikzpicture}[scale=0.5]
\draw[->-] (0,0)--(90:1.5) ;
\draw[->-] (-30:1.5)--(0,0) ;
\draw[->-] (210:1)--(0,0) ;
\draw[->-=0.4] (210:2) .. controls (240:1.5) .. (210:1) ;
\draw[->-=0.9] (210:2) .. controls (180:1.5) .. (210:1) ;
\draw[->-] (210:3)--(210:2) ;
\end{tikzpicture}
\end{array} .
\]
Therefore, the invariance under the $(1,3)$-moves is equivalent to the following bubble-cancellation condition:
\[
\begin{array}{c}
\begin{tikzpicture}[scale=0.5]
\draw[->-] (-90:1.5)--(0,0) ;
\draw[->-] (0,0)--(45:1) ;
\draw[->-] (0,0)--(135:1) ;
\draw[->-] (45:1)--+(135:1) ;
\draw[->-] (135:1)--+(45:1) ;
\draw[->-] (45:1)++(135:1)--+(90:1.5) ;
\end{tikzpicture}
\end{array}
=
\begin{array}{c}
\begin{tikzpicture}[scale=0.5]
\draw[->-] (0,0)--(90:3) ;
\end{tikzpicture}
\end{array} .
\]
This condition means the Frobenius algebra $B$ satisfies $\mu \circ \Delta = \id_B$, i.e. the special condition.

Hence, we conclude that the partition function \eqref{eq:partition_function} constructed by a symmetric special Frobenius algebra $B$ is invariant under Pachner moves, i.e. topological invariant. Also, from the above proof, one can see that to construct a topological invariant partition function we must use a symmetric special Frobenius algebra.

\end{document}